\documentclass[12pt]{article}

\usepackage{epsfig}

\def\bphi{\mbox{\boldmath $\phi$}}

\def\balpha{\mbox{\boldmath $\alpha$}}
\def\bbeta{\mbox{\boldmath $\beta$}}
\def\btau{\mbox{\boldmath $\tau$}}
\def\bnu{\mbox{\boldmath $\nu$}}
\def\bmu{\mbox{\boldmath $\mu$}}
\def\bpsi{\mbox{\boldmath $\psi$}}

\def\bb{\mbox{\bf b}}
\def\bd{\mbox{\bf d}}
\def\be{\mbox{\bf e}}

\def\bs{\mbox{\bf s}}
\def\bt{\mbox{\bf t}}
\def\bc{\mbox{\bf c}}
\def\bu{\mbox{\bf u}}
\def\bv{\mbox{\bf v}}

\def\tfrac#1#2{{\textstyle{{#1}\over{#2}}}}

\def\half{\tfrac{1}{2}}

\def\phivac{\Phi_{\mathrm{\scriptscriptstyle{VAC}}}}

\begin{document}

\begin{titlepage}

\baselineskip 24pt

\begin{center}

{\Large {\bf A First Test of the Framed Standard Model against Experiment}}

\vspace{.5cm}

\baselineskip 14pt

{\large Jos\'e BORDES \footnote{Work supported in part by Spanish MICINN
and FEDER (EC) under grant FPA2011-23596 and
GVPROMETEO2010-056}}\\
jose.m.bordes\,@\,uv.es \\
{\it Departament Fisica Teorica and IFIC, Centro Mixto CSIC, Universitat de 
Valencia, Calle Dr. Moliner 50, E-46100 Burjassot (Valencia), 
Spain}\\
\vspace{.2cm}
{\large CHAN Hong-Mo}\\
h.m.chan\,@\,stfc.ac.uk \\
{\it Rutherford Appleton Laboratory,\\
  Chilton, Didcot, Oxon, OX11 0QX, United Kingdom}\\
\vspace{.2cm}
{\large TSOU Sheung Tsun}\\
tsou\,@\,maths.ox.ac.uk\\
{\it Mathematical Institute, University of Oxford,\\
Radcliffe Observatory Quarter, Woodstock Road, \\
Oxford, OX2 6GG, United Kingdom}

\end{center}

\vspace{.3cm}

\begin{abstract}

The framed standard model (FSM) is obtained from the standard model
by incorporating, as field variables, the frame vectors (vielbeins) 
in internal symmetry space.  It gives the standard Higgs boson and 
3 generations of quarks and leptons as immediate consequences.  It 
gives moreover a fermion mass matrix of the form: $m = m_T \balpha
\balpha^\dagger$, where $\balpha$ is a vector in generation space
independent of the fermion
species and rotating with changing scale, which has already been 
shown to lead, generically, to up-down mixing, neutrino oscillations
and mass hierarchy.  
In this paper, pushing the FSM further, one first derives to 1-loop 
order the RGE for the rotation of $\balpha$, and then applies it to 
fit mass and mixing data as a first test of the model.  With 7 real
adjustable parameters, 18 measured quantities are fitted, most (12) 
to within experimental error or to better than 0.5 percent, and the 
rest (6) not far off.  (A summary of this fit can be found in Table 
\ref{muitable} in the text.)  Two notable features, both generic to FSM, 
not just specific to the fit, are: (i) that a theta-angle of order 
unity in the instanton term in QCD would translate via rotation into 
a Kobayashi-Maskawa phase in the CKM matrix of about the observed 
magnitude ($J \sim 10^{-5}$), (ii) that it would come out 
correctly that 
$m_u<m_d$, despite the fact that $m_t \gg m_b, m_c \gg m_s$.  Of 
the 18 quantities fitted, 12 are deemed independent in the usual
formulation of the standard model.  In fact, the fit gives a total 
of 17 independent parameters of the standard model, but 5 of 
these have not been measured by experiment.

\end{abstract}

\end{titlepage}

\clearpage

\section{Introduction}

We have been suggesting for some time that the standard model 
\cite{databook} be
extended to include the frame vectors \cite{steenrod} in internal 
symmetry space
as field variables (framons), i.e., in addition to the usual gauge
boson and matter fermion fields, resulting in a theory we call
the framed standard model (FSM) \cite{efgt,dfsm}.  The framons 
here are analogous to vierbeins in gravity \cite{ecks} and their 
inclusion 
as fields in particle theory makes it closer in spirit to 
the theory of general relativity and may thus facilitate the 
eventual unification of the two.  As for particle physics itself, 
the immediate attractions for so doing are:

\begin{itemize}

\item It gives the standard Higgs boson, which appears in FSM as 
the framon in the electroweak sector, both a theoretical basis 
and a geometrical significance.

\item It gives to the theory, in addition to local gauge symmetry 
$su(3) \times su(2) \times u(1)$ of the standard model, a global
counterpart, which we may call its ``dual'', $\widetilde{su}(3) 
\times \widetilde{su}(2) \times \tilde{u}(1)$, where the 3-fold
symmetry $\widetilde{su}(3)$ can function as fermion generations
\footnote{while $\widetilde{su}(2)$ is already known to represent 
up-down flavour \cite{tHooft} and $\tilde{u}(1)$ is $B - L$
\cite{prepsm}.}, thus
giving fermion generations a theoretical basis and geometrical
significance as well.

\item It gives a mass matrix for quarks and leptons of the form:
\begin{equation}
m = m_T \balpha \balpha^\dagger,
\label{mfact}
\end{equation}
where $\balpha$, a vector in generation space, is ``universal'' 
(i.e., independent of the fermion species), and rotates with 
changing scale, properties which have been shown \cite{r2m2} to 
lead automatically both to up-down mixing and a hierarchical 
mass spectrum. 

\end{itemize}

Practically, of course, the main attraction of substance is the
last, since fermion mixing and mass hierarchy are two salient 
features imposed on to the standard model by experiment without, 
so far, any theoretical explanation of their origin.  Indeed, it
is the lack of such explanation that accounts for some two-thirds 
of the twenty-odd empirical parameters of the standard model as 
it is presently formulated, and any hint from anywhere towards an 
explanation would be welcome.  Hence, an obvious question to ask 
as a first test of the FSM is whether it can indeed reproduce the 
mass and mixing parameters seen in experiment \cite{databook,ckm,pmns}.  
Previously, the 
FSM has not been sufficiently developed to provide an answer.
The purpose of this paper is to push it far enough to do so, and
the result obtained so far appears to us very positive, as will 
be outlined in the next 2 paragraphs.

At tree-level in the FSM, the vector $\balpha$ appearing in the 
fermion mass matrix (\ref{mfact}) is constant, but corrections by
framon loops make $\balpha$ rotate with changing scale $\mu$.  
Following standard procedure, the RGE for $\balpha$ 
to 1-loop order is then derived.  This equation depends on 5 real 
parameters: 1 coupling $\rho$ and 3 integration constants $a, R_I, 
\theta_I$, plus 1 fudge parameter $k$ representing the dependence 
on $\mu$ of some quantities not yet covered by the present equation.  
However, to make contact with experiment, one needs to supply also the 
coefficients $m_T$ appearing in (\ref{mfact}), one for each fermion 
species, this being the mass of the heaviest state in that species, 
namely $m_t, m_b, m_\tau$ and $m_{\nu_3}$.  Of these the first 3 
have been measured in experiment and are therefore non-adjustable, 
but the last $m_{\nu_3}$, denoting the Dirac mass of the heaviest 
neutrino, is still unknown and has to be treated as a 6th parameter 
of the model.  A 7th parameter is the theta-angle $\theta_{CP}$ 
from the instanton term in the QCD action which, in the FSM, 
translates by rotation \cite{atof2cps,r2m2} into  
the Kobayashi-Maskawa phase of the CKM matrix.  
With these 7 parameters, then, 
the FSM is required to reproduce the mixing matrices of both 
quarks and leptons as well as their masses (except for neutrinos,
the masses of which are likely to be subject to the see-saw mechanism 
\cite{seesaw} and thus 
require more assumptions beyond the basic tenets of the FSM).

For a chosen set of values for these 7 parameters, 23 quantities
have been evaluated, of which 17 are regarded as independent in the 
standard model.  This means that if a good fit to experiment is 
achieved for these quantities, then FSM would have succeeded in 
cutting the number of the relevant standard model
parameters by more than half, from 17 to 7.  
Of the 23 quantities calculated, 18 have been measured in experiment.  
A comparison of the calculated values of these last 18 with data
yields 10 within the stringent experimental errors ($< 1 \sigma$), 
and 2 ($m_\mu, m_e$, for which the minuscule experimental errors 
are beyond the accuracy of our calculation) to within less than 
0.5 percent of the measured values.  Of the other quantities 2 are close
(1.5 and 1.7 $\sigma$ respectively), while the remaining 4, though 
off, are still quite sensible and within striking distances of 
experiment.  We consider that a pretty good score.  The result 
is summarized in Table \ref{muitable} in section 4, at which the 
reader is urged to have a glance before delving into the details 
which follow on how this result is obtained. 

Now, a fermion mass matrix of the factorized form (\ref{mfact}), 
with $\balpha$ rotating with scale, may seem unusual and merits 
perhaps an immediate briefing on how it arises in the FSM.  It is 
basically just a consequence of the framon action being invariant 
under the doubled symmetry $su(3) \times su(2) \times u(1)$ and 
$\widetilde{su}(3) \times \widetilde{su}(2) \times \tilde{u}(1)$ 
already stated.  Because of this, framons in the FSM exist in 
2 types, with the standard Higgs field appearing as the framon 
in the electroweak sector (the ``weak framon''), together with a 
number of new scalar bosons as the ``strong framon'' fields.  As 
usual, it is the Yukawa coupling of the Higgs boson which gives 
the fermion mass matrix, and it is the constraint of the doubled 
invariance on the Yukawa coupling of this ``weak framon'' field
to fermions which gives a tree-level $m$ of the particular form 
(\ref{mfact}).  At tree level, the vector $\balpha$ is a constant, 
but $\balpha$ is coupled to the strong vacuum through the framon 
self-interaction potential, again because of the doubled symmetry.  
Now, the strong vacuum itself is affected by radiative corrections 
via, in particular, the ``strong framons'', and these, being both 
coloured and dual-coloured, will change the orientation of the strong
vacuum 
in $\widetilde{su}(3)$ or generation space and, like other radiative 
correction effects, this change 
would depend on the renormalization scale 
$\mu$.  This forces then the vector $\balpha$ coupled to the strong 
vacuum also to change its orientation in generation space in a 
$\mu$-dependent manner, or in other words, to rotate with changing
$\mu$, as asserted. 

In the following section, we first briefly recall some earlier 
results essential for what follows and collect some tools to be
used for calculating strong framon loops.  Then in  section 3
we shall derive to 1-loop order the RGE of rotation for 
$\balpha$.  And in  section 4, we shall apply 
these RGE to calculate fermion masses and mixing parameters and 
to compare the result with existing data.  In the last section (5)
are remarks about, among other things, the range of validity of 
the derived rotation equation.

\section{Essential Features and Tools}

In this section we recall some properties of the FSM found earlier 
\cite{dfsm,r2m2},
improving the notation wherever merited, while collecting the tools 
needed for the present paper.

The doubled invariance of the FSM under both the (local) gauge 
symmetry $G = su(3) \times su(2) \times u(1)$ and its ``dual'', 
the (global) symmetry $\tilde{G} = \widetilde{su}(3) \times 
\widetilde{su}(2) \times \tilde {u}(1)$, requires that its framon 
fields form a representation of $G \times \tilde{G}$.  Minimality
considerations in the number of fields to be introduced suggest
then that the framons in FSM belong to the representation $({\bf 3}
+ {\bf 2}) \times {\bf 1}$ in $G$ but to $\tilde{\bf 3} \times 
\tilde{\bf 2} \times \tilde{\bf 1}$ in $\tilde{G}$ and that some 
of its components may be taken as dependent on others \cite{efgt,
dfsm}, leaving just:

\begin{itemize}
\item a ``weak framon'' of the form:
\begin{equation}
\balpha \otimes \bphi
\label{wframon}
\end{equation}
where $\balpha$ is a triplet in $\widetilde{su}(3)$, which 
may be taken without loss of generality \cite{dfsm}
as a real unit vector in generation space, but
is constant in space-time, while $\bphi$ is an $su(2)$ doublet but
Lorentz scalar field over space-time which has the same properties 
as, and may thus be identified with, the standard Higgs field;
\item the ``strong framon'':
\begin{equation}
\bbeta \otimes \bphi^{\tilde{a}},\quad  \tilde{a} = \tilde{1}, \tilde{2},
   \tilde{3}
\label{sframon}
\end{equation}
where $\bbeta$ is a doublet of unit length in $\widetilde{su}(2)$ space 
but constant in space-time, while $\bphi^{\tilde{a}}$ are 3 colour
$su(3)$ triplet Lorentz scalar fields over space-time, which when
taken as column vectors give a matrix $\Phi$ transforming by
$su(3)$ transformation from the left but by $\widetilde{su}(3)$
transformations from the right.
\end{itemize}

The doubled invariance under $G \times \tilde{G}$ plus requirements 
of renormalizability restricts the framon self-interaction 
potential to the following form \cite{efgt,dfsm}: 
\begin{eqnarray}
V[\balpha,\bphi,\Phi] & = & - \mu_W |\bphi|^2 + \lambda_W
(|\bphi|^2)^2  \nonumber \\
    & &      - \mu_S \sum_{\tilde{a}} |{\bphi}^{\tilde{a}}|^2
          + \lambda_S \left( \sum_{\tilde{a}}
            |{\bphi}^{\tilde{a}}|^2 \right)^2 
   +   \kappa_S \sum_{\tilde{a},\tilde{b}} 
    |{\bphi}^{\tilde{a}*}\cdot{\bphi}^{\tilde{b}}|^2 \nonumber \\
& &   + \nu_1 |\bphi|^2 \sum_{\tilde{a}} |{\bphi}^{\tilde{a}}|^2
    - \nu_2 |\bphi|^2 |\sum_{\tilde{a}} \alpha^{\tilde{a}}
    {\bphi}^{\tilde{a}}|^2.
\label{VPhi}
\end{eqnarray}
depending on 7 real coupling parameters.  Notice in particular 
the term (with coefficient $\nu_2$) which couples the vector 
$\balpha$ from the weak framon in (\ref{wframon}) to the  
strong framons in (\ref{sframon}).

The vacuum is obtained, as usual, by minimizing the scalar potential, 
in this case the framon potential $V$ in (\ref{VPhi}) .  For $\mu_W, \mu_S$ 
both positive, the case that interests us, the minimum of $V$ 
is degenerate in both the weak and strong sectors.  In the weak sector
the minimum is degenerate in the same way as in the standard model
and little more needs to be said.  It is the degeneracy in the strong 
sector which is at the centre of our present interest.  Here it is 
found \cite{dfsm} that any chosen vacuum in the degenerate set can be 
cast by an appropriate choice of gauge into the following diagonal 
form:
\begin{equation}
\phivac \longrightarrow \zeta_S V_0 
   = \zeta_S \left( \begin{array}{ccc} Q & 0 & 0 \\
                                       0 & Q & 0 \\
                                       0 & 0 & P 
                            \end{array} \right),
\label{Phivac0}
\end{equation}
with:
\begin{eqnarray}
P & = & \sqrt{(1 + 2R)/3}, \\ 
Q & = &\sqrt{(1 - R)/3}, \\ 
R & = & \frac{\zeta_W^2\nu_2}{2 \kappa \zeta_S^2}, 
\label{PQRzetaW}
\end{eqnarray}
where $\zeta_S$ and $\zeta_W$ are the vev's of the strong and 
weak framons respectively, and  $\balpha$, 
which is coupled to the vacuum, takes the value:
\begin{equation}
\balpha \longrightarrow \balpha_0 
   = \left( \begin{array}{c} 0 \\ 0 \\ 1 \end{array} \right).
\label{alpha0}
\end{equation}
Any other vacuum of the degenerate set can be obtained from this, 
but still in the same gauge, by applying a global $\widetilde{su}(3)$ 
transformation, say $A^{-1}$, to (\ref{Phivac0}) from the right 
(and, simultaneously, the same $A$ on $\balpha_0$ from the left). 
The result, of course, will no longer be diagonal, but can be made 
diagonal again by applying an appropriate colour $su(3)$ transform 
from the left.  However, for our present purpose, it is easier to 
keep to just one gauge choice so as to visualize better the change 
of the vacuum with the scale $\mu$.  In other words, we shall pick 
one vacuum of the degenerate set as the reference vacuum, and work 
throughout in the gauge where $\phivac$ and $\balpha$ at that 
reference vacuum take the forms (\ref{Phivac0}) and (\ref{alpha0}) 
above.

We notice that the vacuum (\ref{Phivac0}) has 2 equal eigenvalues.
This is because the original $\widetilde{su}(3)$ in the strong sector
is broken only by the vector $\balpha$ from the weak framon.  There 
is thus still a residual $\widetilde{su}(2)$ symmetry left in the 
system, namely the little group of $\balpha$.  This will be seen to
play a prominent role in constraining the manner that $\balpha$ can
rotate with changing scale, leading to physical consequences of much
interest. 

We are next interested in the quanta of fluctuations of the strong
framon field about its vacuum expectation value (\ref{Phivac0}) which,
by analogy to the Higgs boson in the electroweak sector, we call the 
``strong Higgs'' boson states.  We shall see that it is the exchange 
of these latter states which will give rise to the rotation effect 
we seek.  Again, the analysis can be carried out first around that
particular chosen vacuum labelled by the superscript or subscript 0 
and then extended to other
vacua by an $\widetilde{su}(3)$ transformation.  We choose to express 
these fluctuations as:
\begin{equation}
\phivac^0 + \delta \Phi = \phivac^0 (I + \epsilon S),
\label{Phifluc}
\end{equation}
where $S$ should be taken to be hermitian, 
as any anti-hermitian fluctuation would represent
just a local gauge transformation under local $su(3)$.  We choose for
convenience also to expand $S$ in terms of an orthonormal basis which
is essentially just the unit matrix plus the Gell-Mann matrices, thus:
\begin{equation}
\delta \Phi = \sum_{K} V_K H_K,
\label{expVK}
\end{equation}
where the $H_K$ are then our strong Higgs states, and
\begin{eqnarray}
V_1 & = &  \left( \begin{array}{rrr} 1 & 0 & 0 \\
                                                       0 & 0 & 0 \\
                                                       0 & 0 & 0
                                    \end{array} \right) \nonumber \\
V_2 & = &  \left( \begin{array}{rrr} 0 & 0 & 0 \\
                                                       0 & 1 & 0 \\
                                                       0 & 0 & 0
                                    \end{array} \right) \nonumber \\
V_3 & = &  \left( \begin{array}{rrr} 0 & 0 & 0 \\
                                                       0 & 0 & 0 \\
                                                       0 & 0 & 1
                                    \end{array} \right) \nonumber \\
V_4 & = & \frac{1}{\sqrt{2}} \left( \begin{array}{rrr} 0 & 1 & 0 \\
                                                       1 & 0 & 0 \\
                                                       0 & 0 & 0
                                    \end{array} \right) \nonumber \\
V_5 & = & \frac{i}{\sqrt{2}} \left( \begin{array}{rrr} 0 & -1 & 0 \\
                                                      1 & 0 & 0 \\
                                                       0 & 0 & 0
                                    \end{array} \right) \nonumber \\
V_6 & = & \frac{1}{\sqrt{(P^2+Q^2)}} \left( \begin{array}{rrr}
                                                       0 & 0 & 0 \\
                                                       0 & 0 & Q \\
                                                       0 & P & 0
                                    \end{array} \right) \nonumber \\
V_7 & = & \frac{i}{\sqrt{(P^2+Q^2)}} \left( \begin{array}{rrr} 
                                                       0 & 0 & 0 \\
                                                       0 & 0 & -Q \\
                                                       0 & P & 0
                                    \end{array} \right) \nonumber \\
V_8 & = & \frac{1}{\sqrt{(P^2+Q^2)}} \left( \begin{array}{rrr}
                                                       0 & 0 & Q \\
                                                       0 & 0 & 0 \\
                                                       P & 0 & 0
                                    \end{array} \right) \nonumber \\
V_9 & = & \frac{i}{\sqrt{(P^2+Q^2)}} \left( \begin{array}{rrr}
                                                       0 & 0 & -Q \\
                                                       0 & 0 & 0 \\
                                                       P & 0 & 0
                                    \end{array} \right) \nonumber \\.
\label{VK}
\end{eqnarray}
We shall need these matrices later to calculate strong Higgs loops.

By expanding $V$ in (\ref{VPhi}) further to second order in the fluctuations
of $\Phi$ about its vev, one obtains the tree-level mass matrix 
of the strong Higgs states $H_K$.  This will be given only in the 
last section, for it will not be needed until then.

We turn next to the Yukawa couplings.  Given the usual fermion
fields in the standard model, one can write for the weak framon 
the following Yukawa term:
\begin{eqnarray}
{\cal A}_{\rm YK}^{\rm weak} &=& \sum_{[\tilde{a}] [b]} Y_{[b]} 
\bar{\bpsi}_{[\tilde{a}]}
    \alpha^{\tilde{a}} \bphi \half (1 + \gamma_5) \psi^{[b]}
    + \sum_{[\tilde{a}] [b]} Y'_{[b]} \bar{\bpsi}_{[\tilde{a}]}
    \alpha^{\tilde{a}}\bphi^c \half (1 + \gamma_5) \psi'^{[b]}
    \nonumber \\
&& {} + {\rm h.c.}
\label{wYukawa}
\end{eqnarray}
for leptons (similarly for quarks), 
where the indices $[\tilde{a}]$ and $[b]$ just label copies of
identical $\widetilde{su}(3)$ singlet fields.  As usual, the 
tree-level
quark and lepton mass matrices are obtained by replacing the
Higgs field $\bphi$ by its vev $\zeta_W$, giving rank-1 
matrices, which, by a harmless relabelling of the singlet 
right-handed 
fields, can all be cast into the form (\ref{mfact}) above, with 
$m_T = \zeta_W \rho_W$ where $\rho_W$ is the Yukawa coupling
strength.  The vector $\balpha$ in $\widetilde{su}(3)$ 
generation space is universal because it comes originally 
from the weak framon (\ref{wframon}) and so is independent of 
the fermion field to which the framon is coupled.

Similarly, one can consider a Yukawa coupling of the strong
framon field (\ref{sframon}) but this needs more discussion
and we shall save it for the next section where it will be the
centre of attention for deriving the RGE for the rotation of
$\balpha$.

Once one knows $\balpha$ as a function of the scale $\mu$, then
the conclusions of earlier analyses \cite{r2m2} allow one to 
calculate the masses and state vectors in generation space 
of the various quark and 
lepton states by solving the coupled equations, for the $U$-quarks, 
for example:
\begin{eqnarray}
{\bf t} & = & {\balpha}(\mu=m_t); \nonumber \\
{\bf c} & = & {\bf u} \times {\bf t}; \nonumber \\
{\bf u} & = & \frac{{\balpha}(\mu=m_t) \times {\balpha}(\mu=m_c)}
   {|{\balpha}(\mu=m_t) \times {\balpha}(\mu=m_c)|},
\label{Utriad}
\end{eqnarray}
and:
\begin{eqnarray}
m_t & = & m_U, \nonumber \\
m_c & = & m_U |{\balpha}(\mu=m_c) \cdot{\bf c}|^2, \nonumber \\
m_u & = & m_U |{\balpha}(\mu=m_u)\cdot{\bf u}|^2,
\label{hiermass}
\end{eqnarray}
with $m_U = m_t$ to be taken from experiment.   
The values $m_t, m_c, m_u$ 
are then the masses, and $\bt, \bc, \bu$ (in the absence of 
instantons in QCD, i.e., when $\theta_{CP} = 0$) are the state 
vectors of $t, c, u$.  

However, when there are instantons, as in general there will be, 
and $\theta_{CP} \neq 0$, then CP for quarks will have to be 
redefined by a chiral transformation to eliminate $\theta_{CP}$ 
from the QCD action (i.e., to ``solve the strong CP problem'') so
as to restore CP-invariance to the strong sector \cite{weinberg}.  
(Such chiral
transformations are allowed in the FSM because the quark mass
matrix (\ref{mfact}) has zero eigenvalues \cite{strongcp}.)  The state vectors 
then become:
\begin{eqnarray}
\tilde{{\bf t}} & = & \balpha(\mu = m_t), \nonumber \\
\tilde{{\bf c}} & = & \cos \omega_U \btau(\mu = m_t) - \sin \omega_U
              \bnu(\mu = m_t) e^{-i \theta_{CP}/2}, \nonumber \\ 
\tilde{{\bf u}} & = & \sin \omega_U \btau(\mu = m_t) + \cos \omega_U
              \bnu(\mu = m_t) e^{-i \theta_{CP}/2},
\label{tcutilde}
\end{eqnarray}
where $\btau$ is the unit tangent vector to the trajectory of 
$\balpha$ at $\mu = m_t$, $\bnu = \balpha \times \btau$ is the 
binormal and $\cos \omega_U = \bc\cdot\btau$.  These formulae differ
from those of (\ref{Utriad}) essentially in an extra phase rotation 
$ e^{-i \theta_{CP}/2}$ on the binormal $\bnu$.

Similar considerations hold also for $D$-quarks, which together with the
above then give the CKM matrix as:
\begin{equation}
V_{CKM} = \left( \begin{array}{ccc}
   \tilde{\bf u} \cdot\tilde{\bf d}  &  \tilde{\bf u} \cdot
\tilde{\bf s}  &  \tilde{\bf u} \cdot \tilde{\bf b}  \\
    \tilde{\bf c} \cdot \tilde{\bf d}  &  \tilde{\bf c} \cdot
\tilde{\bf s}  &  \tilde{\bf c} \cdot \tilde{\bf b}  \\
    \tilde{\bf t} \cdot \tilde{\bf d}  &  \tilde{\bf t} \cdot
\tilde{\bf s}  &  \tilde{\bf t} \cdot \tilde{\bf b} 
          \end{array} \right),
\label{CKMtilde}
\end{equation}
which will in general be complex with a nonzero CP-violating KM
phase depending on $\theta_{CP}$, the strong CP problem having
been transmuted by FSM to become the KM phase \cite{atof2cps}. 

Leptons, on the other hand, are not involved, as far as one knows, 
when $\theta_{CP} \neq 0$, so that a solution of the leptonic version 
of the equations (\ref{Utriad}) and (\ref{hiermass}) may already give 
the state vectors, leading to a real PMNS mixing matrix, although 
there is also no compulsion for this being so.  We shall leave the 
question open, to be explored later when fitting data.  The solution 
of the same equations will give also the masses of the charged leptons 
and the ``Dirac masses'' of the neutrinos, but the physical masses of 
the neutrinos will likely be given by some see-saw mechanism and 
cannot be obtained in the present FSM without further assumptions. 

With the above formulae, one will be able to calculate the masses
and mixing parameters for both quarks and leptons (except physical
masses for neutrinos) once one knows the trajectory for $\balpha$.

\section{RGE for the Rotation of $\balpha$}

What we are after in this section is the scale-dependence of the 
vector $\balpha$ which appears in the mass matrix (\ref{mfact}).  
Strictly speaking, $\balpha$, being a vector in $\widetilde{su}(3)$ 
space, carries only global $\tilde{a}$ indices and cannot emit 
or absorb framon or gauge boson quanta which carry local $su(3)$ 
indices, and are thus not subject directly to radiative corrections 
by framon or gauge boson loops.  However, as is seen in (\ref{VPhi}),
$\balpha$ is coupled to the strong vacuum because of the required
double invariance on the framon action, so that if the strong
vacuum, which is subject to radiative corrections, moves with
scale, so must $\balpha$ move as well.  Another way of putting it
is as follows.  The strong vacuum in FSM is degenerate, different
elements in the degenerate set being related by $\widetilde{su}(3)$
transformations.  So if the strong vacuum moves from one element 
to another within this degenerate set, an $\widetilde{su}(3)$
transformation, say $A$, is induced, and $\balpha$, being a vector 
in $\widetilde{su}(3)$ will be transformed by $A$ as well, i.e.,
will rotate.

That being the case, our attention is now directed towards the
strong vacuum in FSM.  Information on how the strong vacuum moves
with scale can be obtained in principle from the RGE of any field 
quantity which depends on the strong vacuum.  Our first choice, 
mainly for historical reasons, fell on a Yukawa-type coupling, 
on which we have had some experience in connection with the mass 
matrix rotation problem. 

However, it is not immediately obvious how to construct a Yukawa
coupling from the strong framon field with the fermion fields that
we have inherited from the standard model, where all left-handed
fields are $su(2)$ doublets and all right-handed fields are $su(2)$ 
singlets.  The strong framon $\Phi$, being a scalar field, couples 
a left-handed to a right-handed fermion field but, being itself 
an $su(2)$ singlet, cannot neutralize the $su(2)$ indices of the 
left-handed fields, as the ordinary $su(2)$-doublet Higgs field 
$\bphi$ does, to make the Yukawa coupling an $su(2)$ invariant.  
One
can nevertheless construct solely from available right-handed fermion 
fields invariant couplings of the following form:
\begin{equation}
{\cal A}_{\rm YK}^{\rm strong} = \sum_{[b]} Z_{[b]} \left[\bar{\psi}^a_R \bphi_a 
   \right]^C_{\tilde{a}} \alpha_Y^{\tilde{a}} \psi_R^{[b]} + {\rm h.c.},
\label{YCJose}
\end{equation}
where $C$ represents charge conjugation, and $\balpha_Y$ is some (it 
can be any) vector in dual colour $\widetilde{su}(3)$ space.  We
notice that the quantity inside the square bracket in the above
expression can be entirely neutral so that the construction is
faintly reminiscent of the manner that the Majorana mass term for
neutrinos is constructed.  But it can also be written
with a $\psi'$
\begin{equation}
\bar{(\psi')}^a \half (1 + \gamma_5) = (\bar{\psi}_R^a)^C
\end{equation}
as:
\begin{equation}
{\cal A}_{\rm YK}^{\rm strong} = \sum_{[b]} Z_{[b]} \bar{\psi'}^a
\bphi^*_a
\cdot \balpha_{Y}
    \frac{1}{2}(1 + \gamma_5) \psi^{[b]} + {\rm h.c.},
\label{sYukawa}
\end{equation}
of the Yukawa type we seek.\footnote{This is superior, we think, 
to an alternative interpretation of the same expression as an 
effective coupling for quarks given before in \cite{dfsm}, the present one
being a renormalizable vertex in the fundamental
fields.}

It has yet to be worked out explicitly which of the fermion fields 
can have couplings of the form (\ref{YCJose}), together with all the details 
of the consequences that these couplings could lead to.  These can
be quite intriguing and intricate and cannot be adequately dealt 
with in the present paper in any case.  But, as far as the effects
of renormalization on the strong vacuum is concerned, which is what 
interests us here, it will not matter, so long as the said coupling 
exists. 

This coupling is particularly convenient for our purpose.  First,
on replacing the strong framon field by its vacuum expectation 
value, it gives a mass matrix ${\bf m}$ depending on the vev of 
$\Phi$, and any change in ${\bf m}$ from renormalization will 
give information on the associated change in the vev of $\Phi$, 
hence also the change in $\balpha$ that we seek.  Secondly, the 
coupling contains also the information on how the strong framons 
$\Phi$, or equivalently the (strong) Higgs states $H_K$, couple
to the fermion states, and allows immediately the calculation
of $H_K$ loops.  We recall that it is the change in orientation of 
$\balpha$ (hence also the strong vacuum) in $\widetilde{su}(3)$
space that is wanted, and the strong framons or the $H_K$, being
the only particles in the theory which carry $\widetilde{su}(3)$ 
indices, are the ones that would affect most directly the change
in orientation.

The information obtained from (\ref{sYukawa}) alone on the rotation 
of $\balpha$ will not be exhaustive, nor necessarily more complete 
than that which can be obtained potentially from the RGE of some 
other quantity.  Nevertheless, it will be seen to be already enough 
for our immediate needs.  An investigation on the RGE of some other 
quantities has started, which might yield further information on 
the scale dependence of $\balpha$.  This should not, however, be in 
contradiction with that obtained here from (\ref{sYukawa}) 
if the theory is self-consistent, but should rather supplement 
whatever might still be missing.

Let us now choose as reference vacuum that which has its $\balpha$ 
the same as the $\balpha_Y$ appearing in (\ref{sYukawa}), and work 
in the gauge where the vacuum expectation values of $\Phi$ and of
$\balpha$ have the forms (\ref{Phivac0}) and (\ref{alpha0}). 

Anticipating the result that the vacuum will change with scale under
renormalization, we shall need to work with a general vacuum related
to the reference vacuum by an $\widetilde{su}(3)$ transformation $A$,
though still in the reference gauge.  For deriving the renormalization
group equations we seek, it is sufficient to take $A$ real, i.e., as a
$3 \times 3$ orthogonal matrix.  Expanding then $\Phi^*$ up to first
order in its fluctuations about the vacuum $\phivac^0 A^{-1}$, thus:
\begin{equation}
\Phi^* = \zeta_S V_0 A^{-1} + \sum_K V^*_K A^{-1} H_K,
\end{equation}
we obtain, on substituting into (\ref{sYukawa}), and introducing for short:
\begin{equation}
{\bf v} = A^{-1} \balpha_0,
\end{equation}
to zeroth order, a (strong fermion) mass term:
\begin{equation}
{\cal A}^{\rm strong}_{\rm mass} 
   = \sum_{a,[b]} \bar{\psi'}^a {\bf m}_{a[b]} \half(1+\gamma_5) 
   \psi^{[b]} + {\rm h.c.},
\label{smassterm}
\end{equation}
with the (strong fermion) mass matrix ${\bf m}$ as:
\begin{equation}
{\bf m}_{a[b]} = \zeta_S (V_0 {\bf v})_a Z_{[b]}, 
\label{smfact}
\end{equation}
and to first order in the fluctuations, a (strong) fermion-Higgs
coupling term:
\begin{equation}
{\cal L}_{Y}^{\rm strong} = \sum_{a,[b]} \bar{\psi'}^a
   \left( \sum_K \Gamma_K H_K \right)_{a[b]} 
   \half(1+\gamma_5) \psi^{[b]} + {\rm h.c.},
\label{sFHterm}
\end{equation}
with
\begin{equation}
\Gamma_K = V^*_K |{\bf v} \rangle \langle Z|.
\label{GammaK}
\end{equation}

Given the (strong) fermion-Higgs couplings $\Gamma_K$, 
the RGE for the (strong)
fermion mass matrix ${\bf m}$ in (\ref{smassterm}) can be derived along
standard lines.  Following \cite{MV,CEL,LWX}, the Feynman diagrams to
1-loop level which give corrections to the Yukawa coupling or the mass 
matrix and the RGE for ${\bf m}$ are listed below:

\vspace*{1cm}

\begin{minipage}{6cm}
\[
(\Gamma_K \Gamma^\dagger_K) {\bf m} \, + \,
{\bf m} \,(\Gamma^\dagger_K \Gamma_K) 
\]
\end{minipage}
\begin{minipage}{9.5cm}
\centering
\includegraphics[scale=0.45]{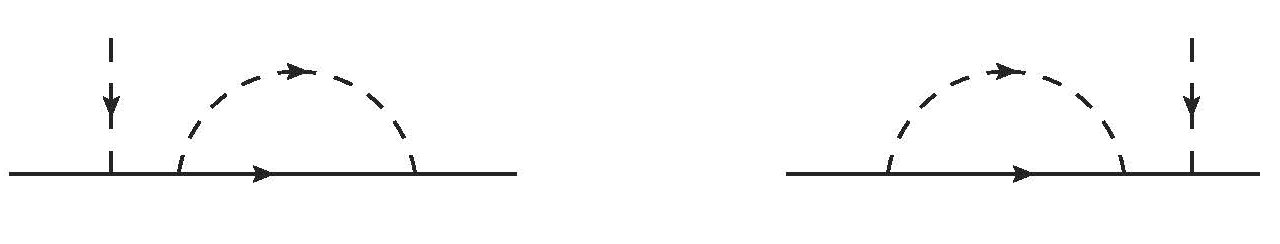}
\end{minipage}

\vspace*{0.5cm}
\begin{minipage}{6cm}
\[
\Gamma_K  \,{\bf m}^\dagger \,\Gamma_K 
\]
\end{minipage}
\begin{minipage}{9cm}
\centering
\includegraphics[scale=0.3]{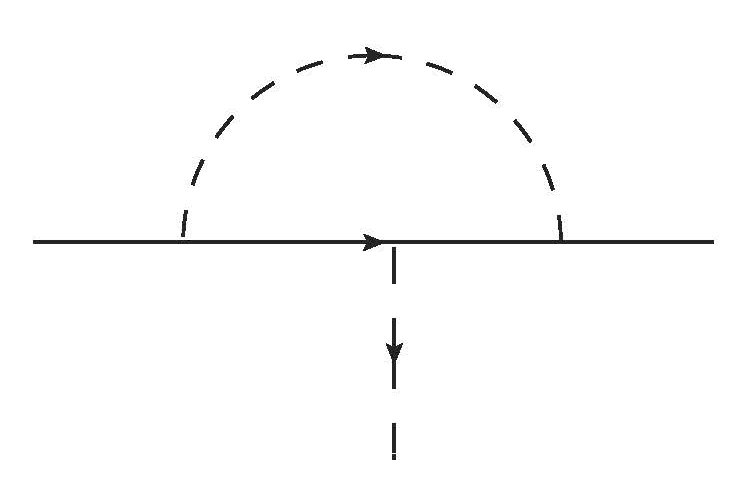}
\end{minipage}

\vspace*{0.5cm}
\begin{minipage}{6cm}
\[
\Gamma_K \,{\tt Tr} \,(\Gamma_K\, {\bf m}^\dagger \, +\, 
{\bf m} \,\Gamma^\dagger_K) 
\]
\end{minipage}
\begin{minipage}{9.2cm}
\centering
\includegraphics[scale=0.35]{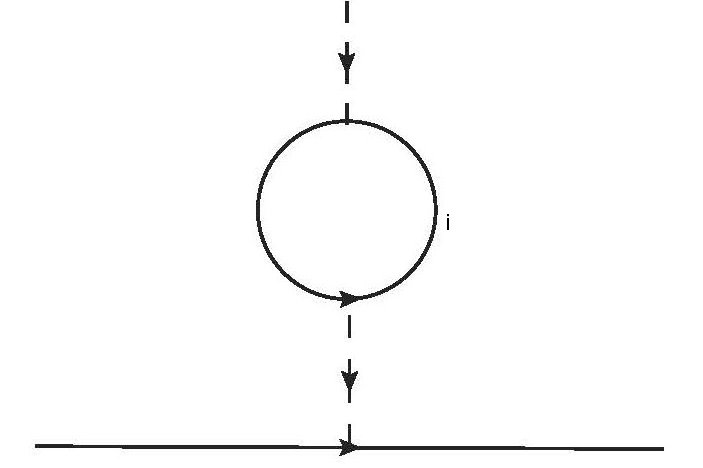}
\end{minipage}

\vspace*{1cm}

Summing the contributions, the RGE of the mass matrix is obtained as:
\begin{eqnarray}
16 \pi^2 \, \frac{d\, {\bf m}}{d\, t}   =   \sum_K \, 
   \left\{ \frac{1}{2} \left[ (\Gamma_K \Gamma^\dagger_K) {\bf m}  + 
   {\bf m} (\Gamma^\dagger_K \Gamma_K) \right] + \right.
\left.
2\Gamma_K  {\bf m}^\dagger \Gamma_K  + 
\Gamma_K {\tt Tr} (\Gamma_K {\bf m}^\dagger + {\bf m} \Gamma^\dagger_K) 
   \right\}
\nonumber \\
\label{RGE}
\end{eqnarray}
where the sum $K$ runs over the whole set of $\Gamma_K$'s in
(\ref{GammaK}) with $V_K$ listed in (\ref{VK}).  This implicitly 
assumes that we are using the RGE at scales above the masses of all 
the (strong) Higgs states $H_K$.  

Next, performing the calculation of the different terms for $\Gamma_K$
in (\ref{GammaK}) and $V_K$ in (\ref{VK}), one has:
\begin{eqnarray}
\Gamma_K \,{\tt Tr} \,(\Gamma_K \,{\bf m}^\dagger & + & {\bf m} \,
   \Gamma^\dagger_K) \, = \,  (\Gamma_K \,\Gamma^\dagger_K) \,{\bf m}
   \, + \, \Gamma_K  \,{\bf m}^\dagger \Gamma_K 
\nonumber \\ \nonumber \\
\sum_K \,  (\Gamma_K \,\Gamma^\dagger_K) \, {\bf m}  & = & 
   \rho_S^2 \, {\rm diag}\, \left(1-v_3^2\,+\,\frac{2 Q^2}{P^2+Q^2}\,
     v_3^2 , \right.
\nonumber \\ \nonumber \\
&&  \quad  \left. 1-v_3^2\,+\,\frac{2 Q^2}{P^2+Q^2}\, v_3^2, \; v_3^2
\,+\, \frac{2 P^2}{P^2\,+\,Q^2} \,(1-v_3^2) \right)\, {\bf m}
\nonumber
\end{eqnarray}

\begin{eqnarray}
{\bf m} \,\sum_K \,  (\Gamma^\dagger_K \,\Gamma_K )  & = &
\rho_S^2 \, \left(\frac{1}{P^2+Q^2}+\frac{3}{P^2+Q^2} 
   (P^2(1-v_3^2) \,+\, Q^2\,v_3^2) \right) {\bf m}
\nonumber \\ \nonumber \\ 
 \sum_K \, \Gamma_K \,{\bf m}^\dagger \, \Gamma_K  & = &\rho_S^2
   \,\left\{{\rm diag} \,\left( 1-v_3^2, \, 1-v_3^2, \, v_3^2 \right) \, 
   + \right.
\nonumber \\ \nonumber \\
&&\quad  +\left.  \frac{2}{P^2+Q^2} \, {\rm diag} \left(P^2\, v_3^2,
   \, P^2\, v_3^2,\, Q^2\,(1- v_3^2),  \right) \right\} \, {\bf m}
\nonumber  \\
\end{eqnarray}
Here
\begin{equation}
\rho_S^2 = \langle {\bf Z}| {\bf Z} \rangle,
\label{rhosq}
\end{equation}
is the Yukawa coupling strength in (\ref{sYukawa}).

Adding the contributions, one has then
\begin{equation}
\frac{d\, {\bf m}}{d\, t}  \,  = \,
{\rm diag} \left( \gamma,  \gamma,  \beta \right)  \, 
   {\bf m}
\label{RGE-explicit}
\end{equation}
where
\begin{equation}
\gamma = \frac{3 \rho_S^2}{16 \pi^2} \left( 2 + \frac{R}{2(2+R)} 
   \right); \ \ \ \  
\beta = \frac{3 \rho_S^2}{16 \pi^2} \left( 2 - \frac{R}{2+R} 
   \right).
\label{betai}
\end{equation}

Notice that the mass matrix ${\bf m}$ defined in (\ref{smfact})
being of rank 1, the matrix 
multiplying ${\bf m}$ on the right-hand side in (\ref{RGE-explicit}) 
is not unique, and may even 
be nondiagonal.  The present diagonal choice, however, 
is particularly convenient for our purpose.  

The equation (\ref{RGE-explicit}) can be rewritten using (\ref{smfact})
in terms of the unit vector ${\bf v}$  as follows:
\begin{equation}
\frac{d}{dt} \ (\rho_S \zeta_S V_0 {\bf v})
   = \left( \begin{array}{ccc} \gamma & & \\ & \gamma & \\
   & & \beta \end{array} \right) (\rho_S \zeta_S V_0 {\bf v})
\label{RGEbfv}
\end{equation}
or else, even more explicitly, in terms of its components as:
\begin{eqnarray}
\frac{\dot{v}_1}{v_1} & = & \gamma - \frac{\dot{Q}}{Q}
   - \left[ \frac{\dot{\zeta}_S}{\zeta_S} + \frac{\dot{\rho_S}}
   {\rho_S} + (QCD) \right], \label{RGEv1} \\
\frac{\dot{v}_2}{v_2} & = & \frac{\dot{v}_1}{v_1}, \label{RGEv2} \\
\frac{\dot{v}_3}{v_3} & = & \beta - \frac{\dot{P}}{P}
   - \left[ \frac{\dot{\zeta}_S}{\zeta_S} + \frac{\dot{\rho_S}}
   {\rho_S} + (QCD) \right] \label{RGEv3} ,
\end{eqnarray}
where we have inserted a term $(QCD)$ in each component for the
possible gluon-loop contribution to the renormalization effect
which, being non-perturbative at low energies and therefore unknown, 
has so far been ignored.  An important point to note here is 
that on the right-hand sides of the equations (\ref{RGEv1})---(\ref{RGEv3}), 
only the terms $\gamma$ or $\beta$ and $\dot{Q}/Q$ or $\dot{P}/{P}$ 
depend on the component.  This means that if it were not for the 
presence of these terms, the vector ${\bf v}$ would not rotate 
with scale at all.  For example, the terms denoted by $(QCD)$ 
coming from gluon-loops will only change the normalization of 
${\bf v}$, not its orientation, and will not by themselves lead 
to any rotation, confirming the earlier statement that only 
strong framon or $H_K$-loops will give rotation directly.  
However, given now that the vector ${\bf v}$ has to remain a unit 
vector, the normalization-changing terms inside
the square brackets, though identical for all components, 
will have an indirect effect on the manner that ${\bf v}$ rotates.
In what follows, we choose  to write the common quantity inside 
the square brackets in (\ref{RGEv1}) and (\ref{RGEv3}) as:
\begin{equation}
\left[\ \cdots\  \right] = -\frac{k}{2} \frac{\dot{R}}{R},
\label{kdef}
\end{equation}
for reasons which will soon be made apparent.  
 
The equation (\ref{RGEv2}) can be integrated to give:
\begin{equation}
v_2 = {\rm constant} \times v_1.
\label{shape1}
\end{equation}
This simple result comes just from the fact that the matrix on the
right of (\ref{RGE-explicit}) has 2 equal eigenvalues, which is
itself the consequence of the residual $\widetilde{su}(2)$ symmetry
already mentioned.  It will be seen to have a significant role in
determining the physical outcome.

Further, for easier manipulation, the equation (\ref{RGEv1}) can 
be replaced by:
\begin{equation}
\beta v_3^2 + \gamma (1 - v_3^2) = -\frac{k}{2} \frac{\dot{R}}{R}
   + \frac{\dot{P}}{P} v_3^2 + \frac{\dot{Q}}{Q} (1 - v_3^2),
\label{normeq1}
\end{equation}
which is derived via (\ref{RGEv1})---(\ref{RGEv3}) from the condition 
that ${\bf v}$ is a unit vector.

Next, the equations (\ref{RGEv3}), (\ref{shape1}) and (\ref{normeq1}) 
on the components of ${\bf v} = A^{-1} \balpha_0$ can be translated 
into conditions on the components of the vector $\balpha = A \balpha_0$ 
of our actual interest using the relations:
\begin{eqnarray}
\alpha^{\tilde{1}} & = & \frac{v_1 v_3}{\sqrt{(v_2)^2 + (v_3)^2}}, 
   \nonumber \\
\alpha^{\tilde{2}} & = & \frac{v_2}{\sqrt{(v_2)^2 + (v_3)^2}}, 
   \nonumber \\
\alpha^{\tilde{3}} & = & v_3.
\label{alphafromv}
\end{eqnarray}

From their definitions in terms of the orthogonal transformation
$A$, one sees that ${\bf v}$ is the third row and $\balpha$ the
third column of the matrix representing $A$ in the basis we are
using, and one would not normally expect any relationship between
them.  The above relations arise here, however, again because of 
the residual symmetry which is the little group of $\balpha_0$.
A rotation about $\balpha_0$ as axis has no effect on ${\bf v}$, 
so that in terms of suitably chosen Euler angles $\theta_i$ and 
the corresponding rotations $R_i$, $i=1,2,3$, we can write
\begin{equation}
A^{-1} = R_1^{-1}R_2^{-1}R_3^{-1}
\end{equation}
and
\begin{equation}
{\bf v} = A^{-1} \balpha_0 = R_1^{-1}R_2^{-1} \balpha_0.
\end{equation}
The RGE's we have derived tell us how ${\bf v}$ rotates with $\mu$,
i.e., how $\theta_1$ and $\theta_2$ change with $\mu$, but give no
$\mu$-dependent constraint on $\theta_3$.  Working out $\theta_1$ 
and $\theta_2$ in terms of ${\bf v}$ and substituting into:
\begin{equation}
\balpha=R_3 R_2 R_1 \balpha_0, 
\end{equation}
then gives the above relationship (\ref{alphafromv}), apart from 
premultiplication by an arbitrary rotation $R_3$ about $\balpha_0$ 
as axis, which however, being $\mu$-independent, is immaterial for 
what follows.
 
Applying then (\ref{alphafromv}) to the equations (\ref{RGEv3}),
(\ref{shape1}) and (\ref{normeq1}) gives the equations in terms 
of $\balpha$ respectively as:
\begin{equation}
\frac{\dot{\alpha}^{\tilde{3}}}{\alpha^{\tilde{3}}} 
   = \beta - \frac{\dot{P}}{P} + \frac{k}{2} \frac{\dot{R}}{R},
\label{RGEa3}
\end{equation} 
\begin{equation}
\frac{\alpha^{\tilde{2}} \alpha^{\tilde{3}}}{\alpha^{\tilde{1}}}
   =  {\rm constant},
\label{shape2}
\end{equation}
and
\begin{equation}
\beta (\alpha^{\tilde{3}})^2 + \gamma [1 - (\alpha^{\tilde{3}})^2]
   = \frac{\dot{P}}{P} (\alpha^{\tilde{3}})^2
     + \frac{\dot{Q}}{Q} [1 - (\alpha^{\tilde{3}})^2]
     - \frac{k}{2} \frac{\dot{R}}{R}.
\label{normeq2}
\end{equation}

Introducing next polar co-ordinates for $\balpha$:
\begin{equation}
\balpha = \left( \begin{array}{lll} \sin \theta \cos \phi \\
                                    \sin \theta \sin \phi \\
                                    \cos \theta \end{array} \right),
\label{alphapol}
\end{equation}
we obtain the RGE for rotation in the form we shall apply
\footnote{The first two of these equations supersede and replace 
the equations (76) and (77) in \cite{dfsm}.  The earlier equations 
were derived before the problem was fully understood and, though 
enough for the then immediate purpose of showing that $\balpha$ 
rotates, contain some errors and omissions in details.  This new 
systematic approach corrects an important sign error, and fills in 
some logical steps as well as the strong framon tadpole
contribution missing in 
the earlier attempt.}:
\begin{eqnarray}
\dot{R} & = & - \frac{3 \rho_S^2}{16 \pi^2} \frac{R(1-R)(1+2R)}{D}
   \left( 4 + \frac{R}{2+R} - \frac{3 R \cos^2 \theta}{2+R} \right)
\label{Rdot} \\
\dot{\theta} & = &  - \frac{3 \rho_s^2}{32 \pi^2} 
   \frac{R \cos \theta \sin \theta}{D} 
   \left( 12 - \frac{6R^2}{2+R} - \frac{3k(1-R)(1+2R)}{2+R} \right)
\label{thetadot} \\ \nonumber
\end{eqnarray}
and
\begin{equation}
\cos \theta \tan \phi = a, 
\label{shape}
\end{equation}
with $a$ constant, and
\begin{equation}
D = R(1+2R) - 3R\cos^2 \theta + k(1-R)(1+2R).
\label{D}
\end{equation}

\section{Fit to Experimental Data}

We shall describe and analyse the fit in some detail because,
given the relative novelty of the FSM mechanism, it is important
not just to get a fit but also to understand how it comes about.

\vspace{.5cm}

The information provided by the renormalization group equations
derived above which govern the rotation trajectory for $\balpha$
divides conveniently into two parts, one (A) specifying the shape 
of the curve, say $\Gamma$, traced out by $\balpha$ on the unit 
sphere as it moves with scale, and the other (B) specifying the 
scale-dependent speed at which it moves along that curve.

\vspace{.5cm}

(A) The shape of $\Gamma$ is given by the equation (\ref{shape}) 
which depends on only 1 parameter, the integration constant $a$.  
Besides, although it was derived above explicitly only at the 
1-loop level, it was seen in (\ref{shape1}) to be a consequence 
of the vacuum being invariant under an $\widetilde{su}(2)$ 
subgroup, namely the little group of $\balpha$, of the global dual 
colour symmetry $\widetilde{su}(3)$, ultimately traceable to the 
same invariance in the framon potential $V$ in (\ref{VPhi}).  One
expects it to hold therefore even at higher-loop levels. 

\begin{figure}
\centering
\includegraphics[height=17cm]{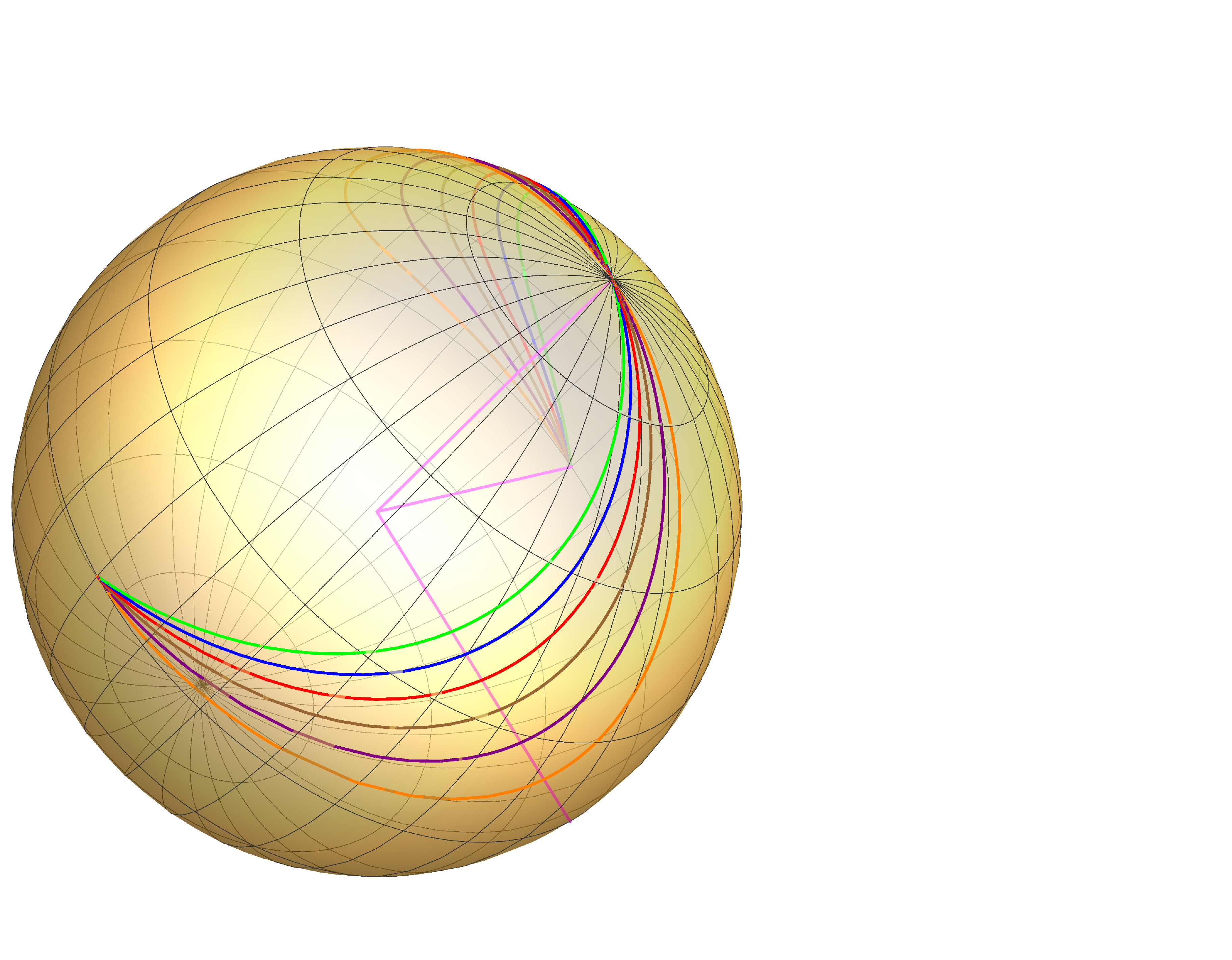}
\caption{The curve $\Gamma$ traced out by the vector $\balpha$
on the unit sphere in generation space for various values of 
the integration constant $a$, decreasing in magnitude from $a=-0.6$ in green to
$a=-0.1$ in orange.}
\label{Gammaplot}
\end{figure}

A plot of $\Gamma$ for several values of $a$ is shown in Figure
\ref{Gammaplot}, where it is seen that it has a sharp bend near
$\theta = \pi/2, \phi = 0$, with the bend becoming sharper
at smaller magnitudes of $a$ \footnote{The sign of $a$, which is
  directly related to the overall sign
of the  geodesic curvature, is immaterial,
as this just fixes the relative orientation of the curve to the sphere.}.
In other words, it has considerable
geodesic curvature somewhere along its length, which, as noted
earlier \cite{features,r2m2}, is essential for the Cabibbo 
angle (i.e., $V_{us} \sim V_{cd}$ in the CKM matrix) to acquire 
the sizeable value it is known experimentally to possess.  It 
is also necessary to give nonzero though still small values to 
the corner elements ($V_{ub}, V_{td}$) of the CKM matrix which
arise only from second-order curvature terms which are
torsion-like \cite{cornerel}.  Furthermore, one notes
that the geodesic curvature changes sign at $\theta = 0$, which 
will be seen later to lead correctly to the result $m_u < m_d$.
This last, of course, is a physically crucial fact, though a 
surprising one, given that in both the 2 heavier generations, 
$m_t \gg m_b, m_c \gg m_s$, and is thus an elusive target for 
model builders.

\vspace{.5cm}

(B) The scale-dependent speed at which $\balpha$ moves along 
the curve $\Gamma$ is prescribed by the equations (\ref{Rdot}) 
and (\ref{thetadot}).  It is much less precisely predicted by 
the FSM than is the shape of $\Gamma$, being known at present 
only to 1-loop, and even at that level depends on 3 parameters, 
namely the Yukawa coupling strength $\rho_S$ and 2 integration 
constants, say $R_I$ and $\theta_I$ at some chosen initial 
scale $\mu_I$.  Even more seriously, it depends also on a quantity 
inside the square brackets in (\ref{RGEv1}) and (\ref{RGEv3}) 
which are unknown and may not even be calculable at present, and 
in the equations (\ref{Rdot}) and (\ref{thetadot}) are just lumped 
together in the unknown function $k(\mu)$.

Still, there are discernible in (\ref{Rdot}) and (\ref{thetadot})
certain features which lead to immediate physical consequences,
having perhaps a more general validity beyond the 1-loop level
at which they are derived.  Thus first, one notices that in 
both equations there is a factor $R$ on the right-hand side in
front, which means that both $\dot{R}$ and $\dot{\theta}$ will
vanish at $R = 0$, or in other words, that there will be a fixed 
point at infinite scale $\mu$ at $R = 0$.  As the scale lowers
from $\infty$, therefore, the change in $\theta$, or rotation,
will accelerate, at least initially for high mass scales.  Now, 
since in the FSM, the masses of the second generation relative 
to the heaviest, are consequences just of rotation and would 
increase for increasing rotation speed, it follows immediately 
from this that $m_c/m_t < m_s/m_b < m_\mu/m_\tau$, simply by
virtue of the fact that $m_t > m_b > m_\tau$, i.e., at least
if the acceleration effect persists that far.

However, there is yet another reason why the noted acceleration 
would persist beyond the immediate vicinity of the fixed point
at $\mu = \infty$.  On the right-hand side of (\ref{thetadot}), there is 
a factor $\sin \theta$ which would seem at first sight to make
$\dot{\theta}$ vanish at $\theta = 0$ so that $\theta$ cannot
then change any further.  Surprisingly, this is not true, at
least for the type of solutions which is of interest to us, for
which the denominator $D$ also vanishes at $\theta = 0$.  This
cancellation is quite intriguing and deserves some examination.
One notices that $D$ being dependent on both $\theta$ and $R$,
its behaviour as $\theta \rightarrow 0$ will depend on the
direction on the $\theta R$-plane in which that limit point is 
approached.  It turns out that for the solutions of interest 
to us, the approach is from near the $\theta$-direction around 
$R = 1$ so that $D$ becomes effectively just $3 \sin^2 \theta$ 
for whatever value of $k$ at the corresponding value of $\mu$.
In other words, $D$ not only cancels off the original zero in
the numerator but even makes $\dot{\theta}$ virtually infinite.
Or that the initial acceleration near the fixed point at $\mu
= \infty$ not only will persist until but will indeed become 
enhanced around the value of $\mu$ at which $\theta$ crosses 
zero.  As to what value of $\mu$ at which this happens, and
why the solutions of interest to us should choose to approach
$\theta = 0$ this way, it will have to be explained later after 
one has had some experience in fitting data.  For the moment,
one will only note that since up-down mixing in the FSM is 
also a consequence of rotation, the fact that rotation will 
continue to accelerate at low scales would immediately imply 
that the mixing angles for leptons will in general be bigger 
than the corresponding ones for quarks, as are actually seen 
in experiment, simply because 
quarks are heavier than leptons in general.

These implications come only from the 2 kinematic factors, $R$
and $D$ in the RGE (\ref{Rdot}) and (\ref{thetadot}), one in
the numerator and one in the denominator, which seem to have 
little to do with the 1-loop approximation from which the RGE
are here derived.  We are thus hopeful that they might persist
even at higher loops. 

\vspace{.5cm}

With these qualitative observations made, we are now ready to
proceed to a quantitative analysis so as to ascertain whether 
the derived renormalization group equations (\ref{shape}), 
(\ref{Rdot}) and (\ref{thetadot}) do actually measure up to 
expectation for the description of experimental data.  These
equations depend on the Yukawa coupling strength $\rho_S$ or 
$c = \rho_S^2/16 \pi^2$, and an at present unknown function
$k(\mu)$.  Clearly, one can do little phenomenologically with
an unknown function.  We shall therefore fudge it and replace
it with a constant $k$, basically assuming that the quantity
inside the square brackets in (\ref{RGEv1}) and (\ref{RGEv3})
has a scale-dependence similar to $\dot{R}/R$, and take $k$
as another parameter to be fitted to experiment.  One can hope
that future renormalization studies of quantities other than
the Yukawa coupling studied here may yield new information on
the function $k(\mu)$.  Here, we can only hope that replacing
it by a constant may not be too drastic, since the quantity in
square brackets, as argued above, are secondary effects 
affecting rotation only indirectly through the normalization 
of the unit vector $\bv$.  

Assuming some values of $c$ and $k$, together with some initial
values for the 2 variables $R, \theta$ at some convenient scale 
$\mu_I$, one can easily integrate numerically the equations to obtain
$R$ and $\theta$ as functions of $\mu$, from which the other
polar angle $\phi$ can be obtained by solving (\ref{shape}) for 
$\Gamma$.  Hence, one has all the information one needs on
the trajectory for $\balpha$.  Supplying then the values of
$m_t, m_b, m_\tau$ from experiment, as listed in Table \ref{mT},
and some assumed values for $m_{\nu_3}$ (the ``Dirac mass'' of 
the heaviest neutrino $\nu_3$) and the theta-angle $\theta_{CP}$
(from the instanton term in the QCD action), one can calculate,
using (\ref{Utriad}), (\ref{hiermass}), and (\ref{CKMtilde}) in 
the Introduction (plus similar formulae for the other fermion 
species), the masses $m_c, m_s, m_\mu, m_{\nu_2}, m_u, m_d, m_e, 
m_{\nu_1}$, as well as the elements of the mixing matrices (CKM 
for quarks and PMNS for leptons).  

\begin{table}
\centering
\begin{tabular}{|l|l|l|}
\hline
& Expt (September 2014) & Input value\\
\hline
$m_t$ & $173.07\pm0.52\pm0.72$ GeV & $173.5$ GeV\\
$m_b$ & $4.18 \pm 0.03$ GeV & $4.18$ GeV \\
$m_\tau$ & $1776.82\pm0.16$ MeV & $1.777$ GeV\\
\hline
\end{tabular}
\caption{Masses of $t$, $b$ and $\tau$ used in our calculation.}
\label{mT}
\end{table}

All in all, then, there are in the scheme 7 adjustable parameters,
namely $a, c, k, R_I, \theta_I, m_{\nu_3}, \theta_{CP}$, of which 
the last 2 double also as part of the physics output.  From these, 
one calculates the 8 lower generation masses, 4 independent numbers
from the CKM matrix and 3 from the PMNS (thus disregarding an unknown
$\delta_{CP}$ for leptons, assumed here to be 0), i.e., altogether 
17 physical quantities (including $m_{\nu_3},\theta_{CP}$) which are 
all regarded as independent parameters in the usual formulation of the
standard model.  In other words, if, by adjusting the 7 parameters, 
one can calculate all these 17 quantities to agree with experiment, 
one would have shown that these 17 independent parameters in the 
standard model can all be replaced by 7 in the FSM.\footnote{By the
standard model here, we mean that in which the now established fact,
that neutrinos have masses and oscillate, is incorporated.  This 
means it will have to carry the Dirac masses of the neutrinos also
as parameters.  Further, we count $\theta_{CP}$ also as a parameter
of the standard model although it is often arbitrarily put to zero.}
 
However, some of these 17 quantities have not yet been measured in, 
or inferred from, experiment.  These include $\theta_{CP}$ from QCD, 
and the Dirac masses $m_{\nu_i}$ of the 3 neutrinos (unless one 
specifies a see-saw mechanism).  Further, because of confinement,
one can infer from experiment only $m_u$ and $m_d$ when run by QCD 
to the GeV scale, but not their values at their own mass scales as 
are needed here.  However, the ratio $m_u/m_d$ should remain constant under
QCD running.  This leaves then, in principle,
only 12 of the original 17 quantities to be fitted.  However, in 
practice, we have to fit more, for although there are in theory,
because of unitarity, only 4 independent parameters in the CKM 
matrix, the requirement of only 4 of the elements $|V_{rs}|$ to be
within experimental errors is no guarantee that the other 5 will 
be within errors too, the unitarity conditions being quadratic 
and some of the elements being much smaller than the others.  For 
this reason, we have to take into account in our fit all the 9 
elements of $|V_{rs}|$ together with the Jarlskog invariant $J$ 
\cite{jarlskog}.  
(The same remarks apply also to the PMNS matrix, but here since 
only 3 mixing angles have been measured experimentally, there is 
no other quantity to fit.)  This then increases the total number 
of measured quantities to 18, which are to be fitted by adjusting 
our 7 parameters. 

We shall proceed in the following manner, which we think will give 
a more transparent test of the FSM scheme than an immediate overall 
fit to all the 18 listed quantities.  We shall first select a subset 
of the listed quantities, the experimental values of which we shall 
supply as input so as to determine, by fitting them, the values 
of the 7 adjustable parameters.  Then with the values for the 7 
parameters so determined, we shall calculate via FSM the values of 
the remaining quantities, which can now be taken as predictions of 
the model to be tested against data, and if they agree, we have a 
direct test of the FSM against experiment. 

We have selected the target quantities to be fitted according 
to the following criteria:
\begin{itemize}
\item that they are sufficient to determine the 7 parameters 
adequately,
\item that they have been measured in experiment to reasonable
accuracy,
\item that they are sufficiently sensitive to the values of the
parameters,
\item that they are strategically placed in $t = \ln \mu^2$ over
the interesting range,
\end{itemize}
and we end up with the following choice: the masses $m_c, m_\mu, 
m_e$ and the elements $|V_{us}|, |V_{ub}|$ of the CKM matrix for
quarks plus the element $|U_{e3}|$  of 
the PMNS matrix for leptons
 (or $\sin^2 2 \theta_{13}$), altogether 6 quantities the measured
values of which as quoted in the PDG are shown in the second column
of the first section of Table \ref{muitable}.  That we
happen to need only these 6 quantities to fix our 7 parameters is 
because of the particular way we have chosen to fit the Cabibbo 
angle, to be explained in the following paragraph, which involves
already by itself the choice of 2 of these parameters.

To fit these 6 quantities with the 7 adjustable parameters is 
then our first task.  We choose to do so by trial and error, 
adjusting judiciously the parameters by hand until we get what 
we consider a decent fit.  We find this much more instructive 
as to how the fit occurs than by, for example, just mechanically 
minimizing $\chi^2$, as it is more traditional perhaps to do.  
In any case, for the present problem, minimising $\chi^2$ would 
be impracticable without introducing some arbitrary weighting 
of the data on the 6 quantities to be fitted, given the great 
disparity in accuracy to which they have been measured in 
experiment, from order $10^{-8}$ for the masses of $\mu$ and 
$e$ to merely about 10 percent for $\sin^2 2 \theta_{13}$.

Besides, in our trial and error fit, we are guided by experience
gathered in previous phenomenological fits to more or less the 
same data \cite{compmec}, so that we know beforehand which of the
7 parameters is likely to affect which of the 6 targeted quantities
most, even though each of the 7 will of course affect all the
quantities concerned to some extent.
First, as already mentioned in (A) at the beginning of this 
section, the large value of the Cabibbo angle $|V_{us}|$ requires 
that the curve $\Gamma$ traced out by the rotating vector $\balpha$ 
to have quite sizeable geodesic curvature  $\kappa_g$  around the 
scale where
the Cabibbo angle is measured.  This means by Figure \ref{Gammaplot} 
that we should choose for $a$ a smallish magnitude.  However, $a$ should 
not be chosen too small; otherwise the geodesic curvature will 
be too concentrated all at one point not to leave enough elsewhere 
to give, for example, a sizeable value also for the solar neutrino 
angle $\theta_{13}$.  A comparison of $\kappa_g$ to previous fits 
\cite{compmec} suggests that a value of $|a| \sim 0.1$ would be 
about right.  Further, to get the maximum benefit of $\kappa_g$
for the Cabibbo angle, one would put $t$ and $b$ at scales fairly
close to the maximum of $\kappa_g$ on $\Gamma$, which means in
practice the parameter $\theta_I$ also fairly close to that same
maximum.  Hence roughly, this takes care of already 2 of the 7 
parameters.

Next, we turn to the masses $m_c, m_\mu, m_e$.  In the rotation
picture, as already noted, these are all consequences of rotation
and would increase with rotation speed, which is in turn governed
by the Yukawa coupling strength $\rho_S$ or $c$.  Hence to fit the
overall sizes of the masses, we can adjust the value of $c$.  The
relative size of $m_c$ and $m_\mu$, on the other hand, is much
conditioned by the proximity of the fixed point at $\mu = \infty$
and $R = 0$, as is explained in (B) at the beginning of this 
section.  Hence, to fit the ratio $m_\mu/m_c$ to data, one can 
adjust  the initial value $R_I$ of $R$.  Finally, to 
fit the value of $m_e/m_\mu$, one can adjust the
fudge parameter $k$ which affects most the speed at low scales.

The rotation speed, of course, affects also the overall size of 
the 2 mixing angles $|V_{us}|$ and $|V_{ub}|$, but not by much 
their relative sizes.  These can be adjusted by adjusting the
parameter $\theta_{CP}$, the theta-angle from the instanton term
in the QCD Lagrangian which in FSM, we recall, is related to the 
Kobayashi-Maskawa phase in the CKM matrix.  

And with this, we have control now on 6 of the adjustable parameters, 
namely $a, c, R_I, \theta_I, \theta_{CP}, k$, leaving on our list
only the Dirac mass of the heaviest neutrino $m_{\nu_3}$ which
affects only quantities involving neutrinos.  Hence, forgetting 
those for the moment, we have a good idea how to adjust these 6
parameters to fit the  5 chosen quantities not involving
neutrinos, namely $m_c, m_\mu, m_e, |V_{us}|, |V_{ub}|$, and
after a few tries, it is not hard to end up with a decent fit.
Then, having got this, it is relatively easy to vary $m_{\nu_3}$
until we have a fit for $\sin^2 2 \theta_{13}$ also.  Our best
result is shown in the third column of the first section of 
Table \ref{muitable}, which has been obtained for the following 
values of our 7 parameters:
$$
a = -0.1, \ \ c = 0.12,\ \ \theta_{CP} = 1.78, \ \ k = 2.05,$$
\begin{equation}
R_I = 0.01, \ \ \theta_I = - 1.33, \ \ m_{\nu_3} = 29.5\ {\rm MeV},
\label{params}
\end{equation}
where the two integration constants (initial values) $R_I$ 
and $\theta_I$ have been taken at $\mu = 250\ {\rm GeV}$ 
for convenience.\footnote{The negative value we quote for the polar 
angle $\theta_I$ may seem surprising, but it is actually no more 
than a convention (see Appendix) adopted for convenience in our 
numerical calculation.}  One notes in Table \ref{muitable} that 
the 4 quantities $m_c, |V_{us}|, |V_{ub}|, \sin^2 2 \theta_{13}$ 
have all been fitted to within present experimental errors, 
while $m_\mu$ and $m_e$, the experimental errors for which are 
so small as to be inappropriate for us to chase, have been 
fitted to, respectively, 0.2 and 0.4 percent.  This is more 
than enough accuracy for our present purpose.

\begin{table}
\centering
\begin{tabular}{|l|l|l|l|l|}
\hline
& Expt (June 2014) & FSM Calc & Agree to & Control Calc\\
\hline
&&&& \\
{\sl INPUT} &&&&\\
$m_c$ & $1.275 \pm 0.025$ GeV & $1.275$ GeV & $< 1 \sigma$&$1.2755$ GeV\\
$m_\mu$ & $0.10566$ GeV & $0.1054$ GeV & $0.2 \%$ & $0.1056$ GeV\\
$m_e$ & $0.511$ MeV &$0.513$ MeV & $0.4 \%$ &$0.518$ MeV\\
$|V_{us}|$ & $0.22534 \pm  0.00065$ & $0.22493$ & $< 1 \sigma$ &$0.22468$\\
$|V_{ub}|$ & $0.00351^{+0.00015}_{-0.00014}$& $0.00346$ & $< 1 \sigma$&$0.00346$ \\
$\sin^2 2\theta_{13}$ & $0.095 \pm 0.010$ & $0.101$ &$< 1 \sigma$ &$0.102$\\
\hline
&&&& \\
{\sl OUTPUT} &&&&\\
$m_s$ & $0.095 \pm 0.005$ GeV & $0.169$ GeV & QCD &$0.170$ GeV \\
& (at 2 GeV) &(at $m_s$) &running& \\
$m_u/m_d$ & $0.38$---$0.58$ & $0.56$ &  $< 1 \sigma$&$0.56$ \\
$|V_{ud}|$ &$0.97427 \pm 0.00015$ & $0.97437$ & $< 1 \sigma$&$0.97443$ \\
$|V_{cs}|$ &$0.97344\pm0.00016$ & $0.97350$ & $< 1 \sigma$&$0.97356$ \\
$|V_{tb}|$ &$0.999146^{+0.000021}_{-0.000046}$ & $0.99907$ &$1.65
\sigma$&$0.999075$ \\
$|V_{cd}|$ &$0.22520 \pm 0.00065$ & $0.22462$ & $< 1 \sigma$ &$0.22437$\\
$|V_{cb}|$ & $0.0412^{+0.0011}_{-0.0005}$ & $0.0429$ & $1.55 \sigma$&
$0.0429$ \\
$|V_{ts}|$ & $0.0404^{+0.0011}_{-0.0004}$ & $0.0413$ &$< 1 \sigma$& 
$0.0412$\\  
$|V_{td}|$ & $0.00867^{+0.00029}_{-0.00031}$ & $0.01223$ & 41 \% & $0.01221$\\
$|J|$ & $\left(2.96^{+0,20}_{-0.16} \right) \times 10^{-5}$ & $2.35
\times 10^{-5}$ & 20 \% &$2.34\times 10^{-5}$ \\
$\sin^2 2\theta_{12}$ & $0.857 \pm 0.024$ & $0.841$ &  $< 1 \sigma$& $0.840$\\ 
$\sin^2 2\theta_{23}$ & $>0.95$ & $0.89$ & $> 6 \%$ &$0.89$\\
\hline 
\end{tabular}
\caption{Calculated fermion masses and mixing parameters compared with
experiment} 
\label{muitable}
\end{table}

Note that the functional form for the trajectory for $\balpha$
(i.e., both the shape of the curve $\Gamma$ it traces out on the
unit sphere and the variable speed at which it moves along that 
curve) having already been prescribed, it is not at all obvious 
that the 6 targeted quantities can be so fitted with the given 
7 parameters.  That it can indeed be done to the accuracy seen
in Table 
\ref{muitable} constitutes already a quite nontrivial test.

The real test for the model, however, arises when the fitted
values of the parameters in (\ref{params}) are used to evaluate
the 12 other measured quantities which have not been selected by 
us as targets for fitting.   As far as the standard model is
concerned, 6 of these are independent of one another and of the 
6 quantities which have been fitted, and so, in principle, can
have any value they choose, and if the RGE derived from the FSM
can reproduce their values, even approximately, it would be a
much welcome success.  They cannot be claimed as predictions, 
for their empirical values are known, and we can and did monitor 
their theoretical values produced by our scheme as we vary our 
parameters to perform the fit.   But there is little we can do
to influence their values since the parameters have to be 
chosen to fit the 6 selected quantities as we have done above.
In other words, once the above fitting procedure is adopted,
the values of the output quantities are entirely at the mercy
of the rotation equations and out of our control altogether. 
Their agreement or otherwise with experiment would thus be a
genuine test for the FSM's validity.

The output quantities of our fit with (\ref{params}) are shown
in the third column and compared with their present experimental
values in the second column of the second section of Table
\ref{muitable}.  We note that of the 12 numbers shown, 6 are 
within experimental error or $1 \sigma$ or else ($m_\mu, m_e$)
within 0.5 percent of the accurate measured values, while 2 are 
within $\sim 1.5 \sigma$.  Of the remaining 4, 1 ($m_s$) can only 
be roughly compared with experiment, because of QCD running as 
already explained, and it does so compare quite reasonably.  The
other 3: $|V_{td}|, J, \sin^2 2 \theta_{23}$, are all outside
the stringent experimental errors, but still not outrageously
so.  Besides, $|V_{td}|$ and $J$ both being small and therefore
delicate to reproduce, obtaining them with the right order of
magnitude as they are here is already no mean task.  We note in
particular the interesting output for $m_u/m_d$ which not only
confirms the elusive but crucial empirical fact that $m_u < m_d$
but even gives the ratio within the present PDG quoted error.
That the ratio should agree numerically may perhaps be a little 
fortuitous, but that $m_u$ should come out smaller than $m_d$
appears generic to the FSM, which is an important point to 
which we shall shortly return.

To us, this is about as good a result as one can hope to get 
with the approximate RGE (\ref{shape}), (\ref{Rdot}) and
(\ref{thetadot}) derived merely at 1-loop and especially with 
a fudge on $k$, replacing that function of scale by a constant.
Indeed, we do not mean, by carrying as many significant figures
as we do in the numbers cited in Table \ref{muitable}, that we
believe the set-up, as it is, is actually correct to this sort
of accuracy.  We do so merely as a test, to see to what sort 
of accuracy that an FSM fit is capable of.  Besides, with our 
trial and error approach to fitting, we cannot claim that the 
present fit is the best fit even in the vicinity in parameter 
space of the present one.  There may also be even better fits
in other parts of the parameter space, e.g.\ when $R < 0$, that 
we have not yet sufficiently explored.  But one can reasonably 
claim, we think, that this fit has demonstrated that the FSM
is capable of giving a very sensible description of the mass 
and mixing data, as one has hoped.

To check our numerical result, a second calculation is done by
another program in another language using the same parameters
(\ref{params}).  This gives the result shown in the last column 
of Table \ref{muitable}.  Besides confirming the result of the
earlier calculation displayed in the third column, it provides 
an estimate for the numerical accuracy of our results, which 
otherwise would not be easy to obtain.  Some numerical details 
of the calculation is given in the Appendix so that interested 
readers can do spot checks on our numbers if they so wish.

The fit gives in addition the following values for the 5 other
standard model parameters which, not being measured, cannot as 
yet be checked against experiment:
$$
\theta_{CP} = 1.78, \ \ m_u(\mu=m_u) = 0.22\ {\rm MeV}\
[{\rm or}\ m_d(\mu=m_d) = 0.39\ {\rm MeV}],$$
\begin{equation}
m_{\nu_3} = 29.5\ {\rm MeV}, \ \ m_{\nu_2} = 16.8\ {\rm MeV},
\ \ m_{\nu_1} =  1.4\ {\rm MeV}.
\label{5others}
\end{equation}

\vspace*{5mm}

\begin{figure}
\centering
\includegraphics[height=7.8cm]{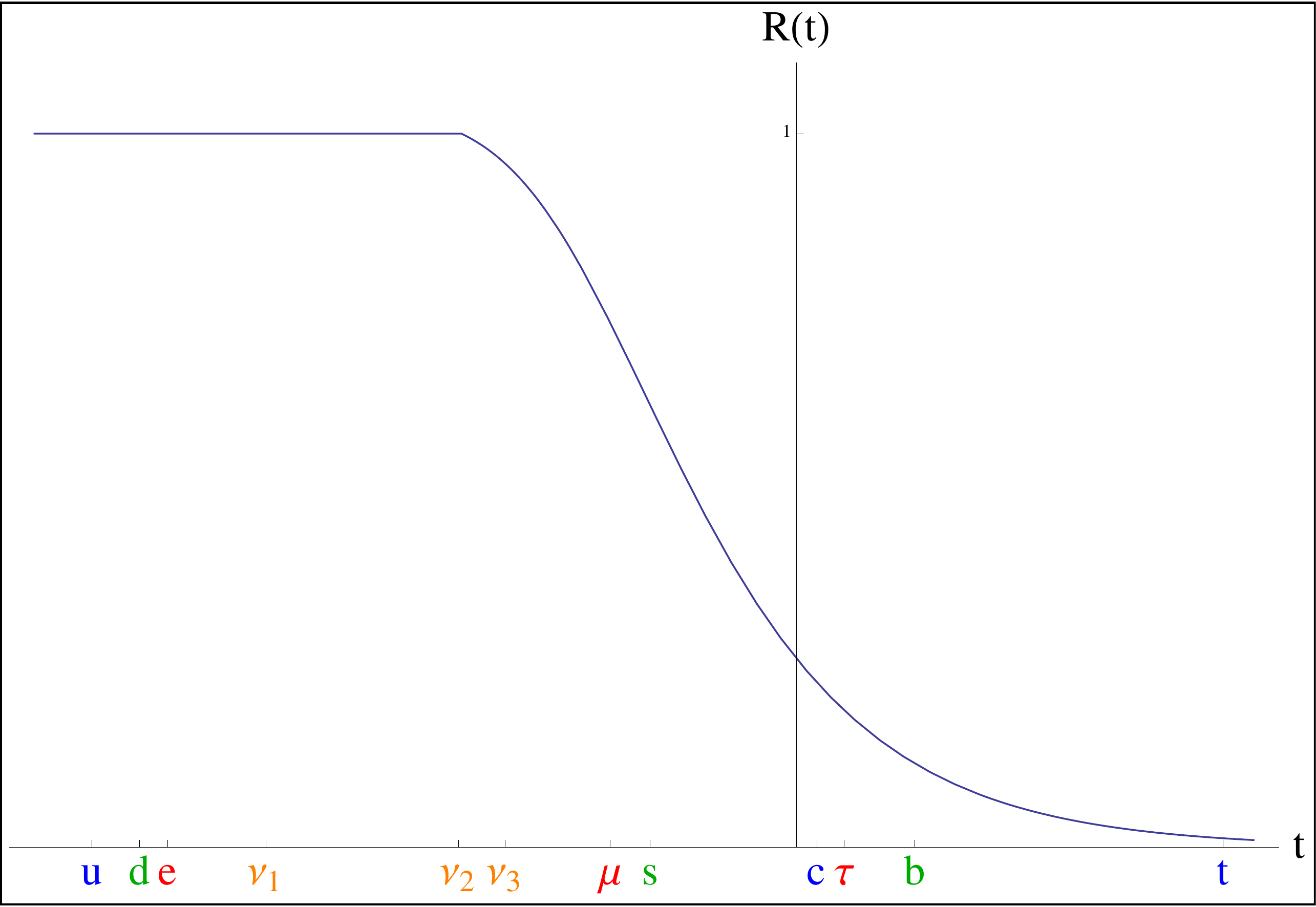}
\caption{Solution of the RGE (\ref{Rdot}) 
for $R$  as a function of $t= \log \mu^2$, where $\mu$ is the scale in
GeV,  
obtained with parameters given as in (\ref{params}).}
\label{Rplot}
\end{figure}

\begin{figure}
\centering
\includegraphics[height=7.5cm]{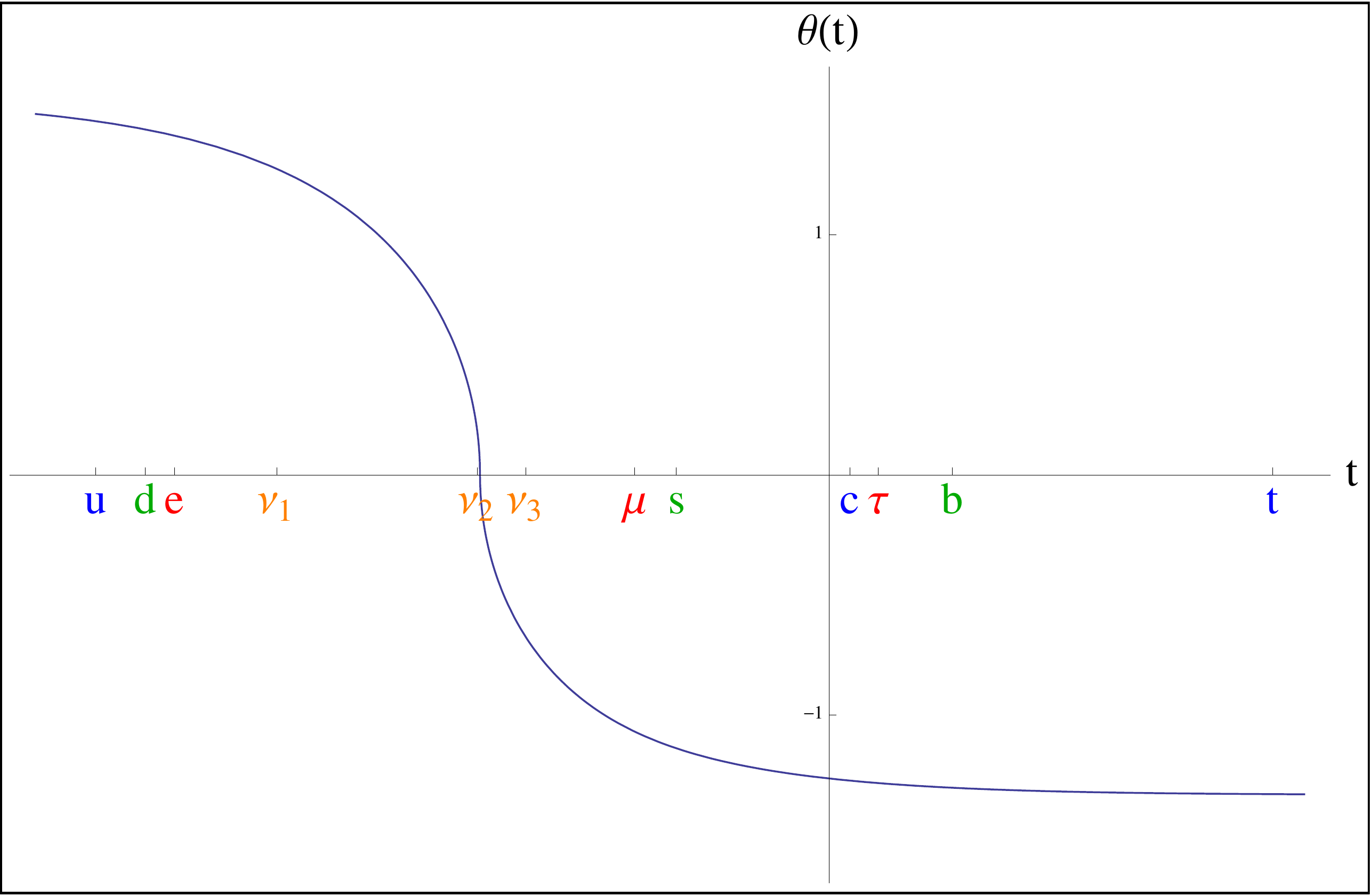}
\caption{Solution of the RGE (\ref{thetadot})
$\theta$  as a function of $t= \log \mu^2$, where $\mu$ is the scale
in GeV,
obtained with parameters given as in (\ref{params}).}
\label{thetaplot}
\end{figure}

\begin{figure}
\centering
\includegraphics[height=17cm]{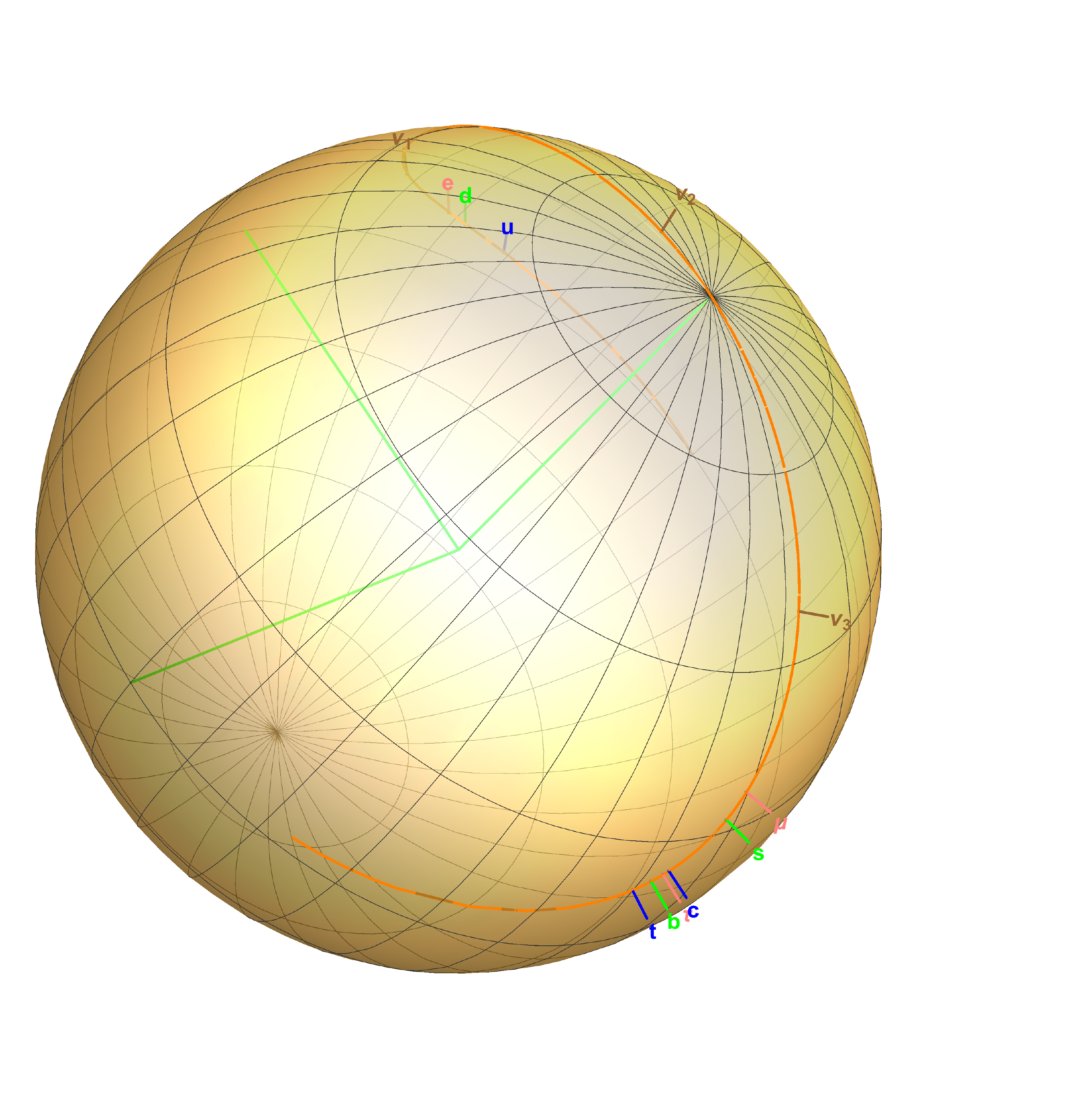}
\caption{The trajectory for $\balpha$ on the unit sphere in
generation space obtained from the parameter values given
in (\ref{params}), showing the locations on the trajectory
where the various quarks and leptons are placed: high scales in front.}
\label{muiplot1}
\end{figure}

\begin{figure}
\centering
\includegraphics[height=17cm]{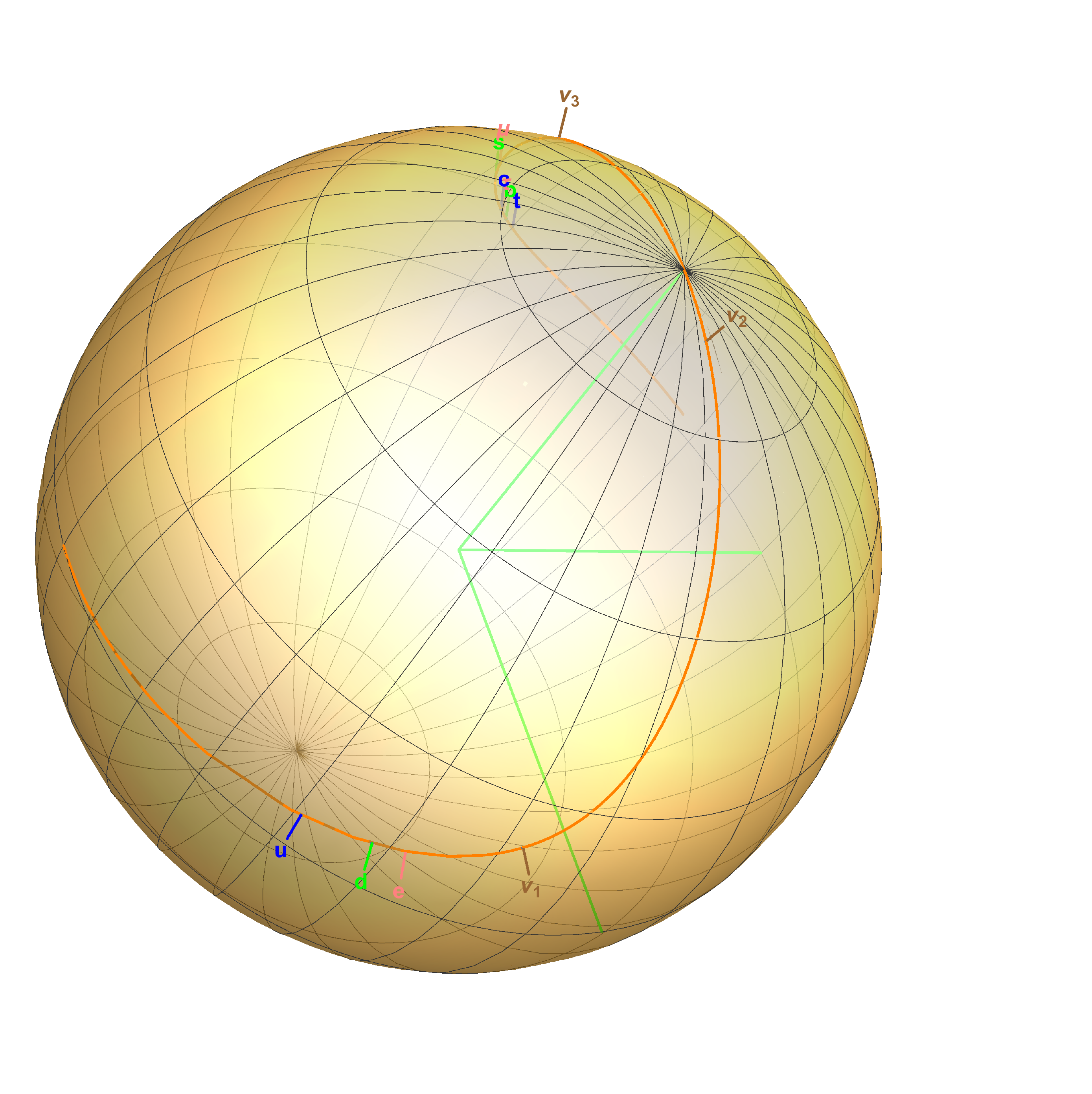}
\caption{The trajectory for $\balpha$ on the unit sphere in
generation space obtained from the parameter values given
in (\ref{params}), showing the locations on the trajectory
where the various quarks and leptons are placed: low scales in front.}
\label{muiplot2}
\end{figure}

\begin{figure}
\centering
\includegraphics[height=15cm]{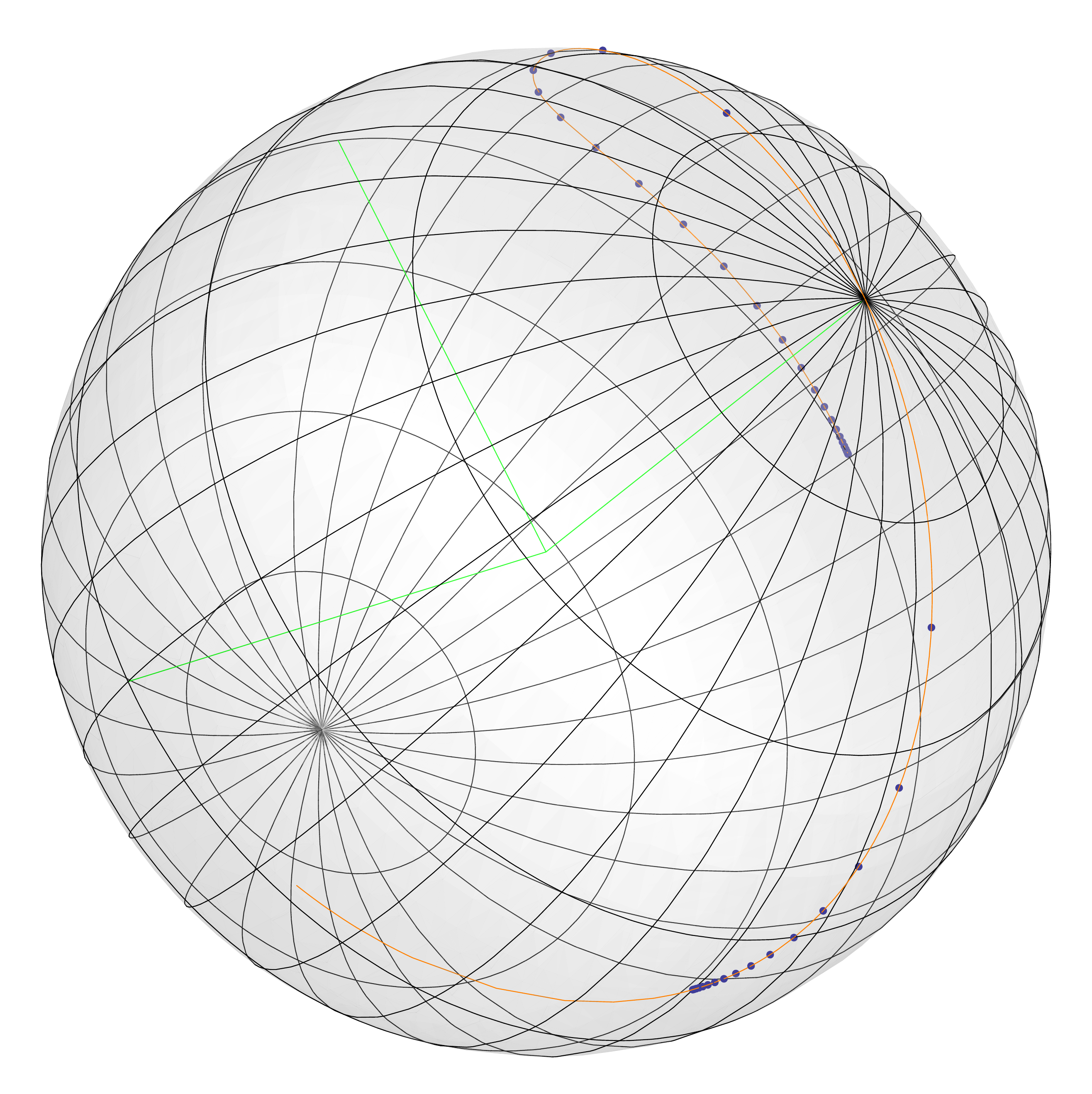}
\caption{The trajectory for $\balpha$ on the unit sphere in
generation space obtained from the parameter values given
in (\ref{params}), showing the distance on $\Gamma$ travelled 
by $\balpha$ for every half-decade decrease in scale $\mu$. The high
scale region is shown in front.}
\label{muiplot3}
\end{figure}

The curve traced out by the rotating $\balpha$ for the value of
$a$ listed in (\ref{params}) is shown already as the curve in
orange in Figure \ref{Gammaplot}.  The solutions of (\ref{Rdot})
and (\ref{thetadot}) for $R$ and $\theta$ as functions of the 
scale $\mu$ are shown in Figures \ref{Rplot} and \ref{thetaplot}.  
The actual
trajectory for $\balpha$ on the unit sphere is given in Figures 
\ref{muiplot1}, \ref{muiplot2} and \ref{muiplot3}, 
and from these, one gets a 
clear visual picture of how, qualitatively, most of the results 
in Table \ref{muitable} come about.

From the fixed point at $R = 0$ and $\mu = \infty$, one sees 
from Figures \ref{thetaplot}  and \ref{muiplot3} that
rotation starts slowly, but accelerates for decreasing 
$\mu$ as it is expected that it should.  As explained in (B)
above, this would account, by the leakage mechanism, for the 
experimental fact that $m_c/m_t < m_s/m_b < m_\mu/m_\tau$.
Comparing Figures \ref{muiplot1} and 
\ref{muiplot3}, one sees
that in the region where these 6 fermions are placed, rotation
is still rather slow, so that to get a sizeable Cabibbo angle,
$\Gamma$ would need to have large geodesic curvature in this
region, and it is seen that it does, which is occasioned by
us choosing in (\ref{params}) a small magnitude for $a$ (0.1) and
a value for $\theta_I$ ($-1.33$) close to the maximum for the
geodesic curvature.  As the scale lowers further, acceleration
is seen to continue with, in fact, the rotation speed becoming 
very large around $t \sim -3.5$ or $\mu \sim 17\ {\rm MeV}$.  
Now such a rapid rise cannot be ascribed to the said fixed 
point at $R = 0$ alone, but arises from another source in the
RGE, namely from the denominator $D$ in (\ref{thetadot}) as
explained in (B) above.  One notes from Figures \ref{Rplot}
and \ref{thetaplot}
that, at these low values of $\mu$, $R$ is already very near  
its asymptotic value $1$, so that the situation is very close
to that described in (B) above.  It gives thus fast rotation 
at the sort of scales just where, for example, $m_{\nu_3}$ is 
placed, and where mixing angles of leptons are evaluated.  It 
follows then, as already explained in (B), that mixing angles 
for leptons are generally larger than those for quarks, even 
though the geodesic curvature there at the mass scales for 
leptons is already far from maximum.

One sees therefore that from the qualitative features of the
trajectory for $\balpha$, one can already envisage an outline
for the mass and mixing patterns vaguely similar to what is 
experimentally observed.  But that a detailed fit of the data
to the accuracy actually achieved in Table \ref{muitable} is
surprising and cannot at first be anticipated.  

As the scale lowers further still, the effect of both the fixed
point at $R = 0$ and the denominator $D$ dies down and rotation 
will slow down.  But there is yet another twist in the tale from the
tail of $\Gamma$ with interesting consequences.  One is now at
scales of order MeV where masses of the lowest generation quarks
occur.  Now according to (\ref{hiermass}) above, the mass of 
the $u$ and $d$ quarks in FSM are to be given respectively by 
solution of the equations:
\begin{equation}
|\langle {\bf u}|\balpha(\mu) \rangle|^2 = \mu, \ \ \ 
|\langle {\bf d}|\balpha(\mu) \rangle|^2 = \mu,
\label{udmass}
\end{equation}
where ${\bf u}$ the state vector of $u$ is of 
course orthogonal to ${\bf t}$ and ${\bf c}$, 
the state vectors of $t$ and $c$.  
Similarly for the triad ${\bf b}, {\bf s}, {\bf d}$.  The
masses of $u$ ($d$) being only of order MeV, this means that 
one has an approximate solution for $m_u$($m_d$) whenever the
vector $\balpha$ crosses the ${\bf t}{\bf c}$-plane (${\bf b}
{\bf s}$-plane).  Given the ordering of the masses of $t,b$ 
and that, as noted before, $m_c/m_t < m_s/m_b$, the picture is
as shown in Figure \ref{udmasspic}.  It is thus clear that in
the MeV region where the geodesic curvature has the opposite 
sign to that in the high scale region, the vector $\balpha$
must cross the ${\bf b}{\bf s}$-plane before (i.e.\ at a higher 
scale than) the ${\bf t}{\bf c}$-plane.  In other words, $m_d$ 
must be larger than $m_u$, as experiment wants.\footnote{One
has chosen here the second (lower) solution for $u$ and $d$ as
one did also for $e$ in (\ref{muitable}).  For more elucidation
on this point, see Appendix.}

\begin{figure}
\centering
\includegraphics[height=18cm]{intersectud2.pdf}
\caption{Figure illustrating the reason why $m_u < m_d$ in Table
\ref{muitable}.}
\label{udmasspic}
\end{figure}

That $m_u < m_d$ is of course a crucial empirical fact, without 
which the proton would be unstable and we ourselves would not
exist.  It is, however, surprising from the theoretical point 
of view, given that in the two heavier generations, the ordering
of masses is the other way round, namely $m_t \gg m_b, m_c \gg m_s$.
This is particularly so from the point of view of rotating mass
matrix (R2M2) schemes \cite{r2m2}, 
of which the FSM is one, in which masses
for the lower generations are supposed to come from ``leakage''
due to rotation from the heavier states; so why should the up 
states leak so much less to the $u$ than the down states to the 
$d$?  It was a question we kept asking since we first began with
the rotation mechanism, but failed to answer.  Intriguingly, it 
is now seen to turn upon the fact that the geodesic curvature 
$\kappa_g$ of  the curve $\Gamma$ changes sign when it passes 
through the pole $\theta = 0$.  The sign-change occurs whatever 
is the value of the integration constant $a$ upon which $\Gamma$ 
depends.  And this shape of $\Gamma$ in the FSM, we recall, is
the consequence of a symmetry in the framon potential $V$ 
in (\ref{VPhi}), which is itself a consequence of the intrinsic 
symmetry of the framed theory.  It is thus a property intrinsic
and peculiar to the FSM, which is hard to foretell or imagine 
otherwise.  Phenomenologically, extrapolating from the higher 
scale region where most of the data lie, one would tend, as 
we did \cite{compmec}, to assume the same sign for $\kappa_g$ 
throughout, and come to the wrong conclusion.  

As already mentioned, no estimate of the physical masses of
neutrinos can be made in the above FSM fit without additional 
assumptions on the see-saw mechanism.  If we now assume the 
simplest, namely that there is only one right-handed
neutrino with mass 
$m_R$, then the physical neutrino masses are given in terms of 
the Dirac masses just as $m^{\rm ph}_{\nu_i} = m_{\nu_i}^2/m_R$ 
\cite{seesaw}.  
Hence, choosing $m_R \sim 17430\ {\rm TeV}$ to fit the experimental 
value:
\begin{equation}
(m^{\rm ph}_{\nu_3})^2 - (m^{\rm ph}_{\nu_2})^2 = (2.23^{+0.12}_{-0.08})
   \times 10^{-3} \  {\rm eV}^2,
\label{Del32}
\end{equation}
one has:
\begin{equation}
m^{\rm ph}_{\nu_3} \sim  0.050 \  eV, \ \ \  m^{\rm ph}_{\nu_2} \sim  
   0.016 \ {\rm eV}, \ \ \  m^{\rm ph}_{\nu_1} \sim 0.0001 \ {\rm eV}.
\label{bfmnui}
\end{equation}
Apart from satisfying the, at present, very loose experimental 
bounds on these individual mass values, this gives:
\begin{equation}
(m^{\rm ph}_{\nu_2})^2 - (m^{\rm ph}_{\nu_1})^2 \sim 2.6 \times 10^{-4}
   \ {\rm eV}^2,
\label{Del21}
\end{equation}
which is at least of the same order as, though some 3 times
bigger than, the experimental value of:
\begin{equation}
(m^{\rm ph}_{\nu_2})^2 - (m^{\rm ph}_{\nu_1})^2 = (7.5 \pm 0.20) \times
   10^{-5} \ {\rm eV}^2.
\label{Del21exp}
\end{equation}
The discrepancy, however, is not as bad as it looks, for this
quantity depends on the rotation angle raised to the 8th power,
and the present discrepancy corresponds to an error of only 17
percent in the rotation angle, which is not exorbitant in the
scale region where the $\nu_2$ finds itself in the fit, namely
near $\theta =0$ where, as seen in Figures  
\ref{thetaplot} and \ref{muiplot3} the
rotation speed is very large.

\section{Remarks}

\hspace*{\parindent}[I] The RGE for the rotation of $\balpha$ 
derived in Section 3 
from renormalization by framon loops was applied down to scales 
as low as the electron mass $\sim {\rm MeV}$.  This presupposes
that there are framons of such low masses to drive rotation at
such scales, for it is generally believed that particles of high
mass cannot give renormalization effects at scales much lower 
than their own mass scales \cite{appleq}.  One would like to check, therefore, 
whether in the FSM scheme there are indeed framons of such low 
masses.

Expanding the framon potential (\ref{VPhi}) to second order in
fluctuations about its vev, one obtains to tree-level the mass 
matrix of the framon-Higgs states (i.e., the standard Higgs $h_W$
together with the 9 strong Higgs states $H_K$) as:
\begin{equation}
M_{H}=
\left( \begin{array}{cccc}
4\lambda_{W}\zeta_{W}^{2} & 2\zeta_{W}\zeta_{S}(\nu_{1}-\nu_{2})
\sqrt{\frac{1+2R}{3}} &
2\sqrt{2}\zeta_{W}\zeta_{S}\nu_{1}\sqrt{\frac{1-R}{3}} & 0 \\
\ast & 4(\kappa_{S}+\lambda_{S})\zeta_{S}^{2}\left(\frac{1+2R}{3}\right) & 
4\sqrt{2}\lambda_{S}\zeta_{S}^{2}\frac{\sqrt{(1+2R)(1-R)}}{3} & 0 \\
\ast & \ast &
4(\kappa_{S}+2\lambda_{S})\zeta_{S}^{2}\left(\frac{1-R}{3}
\right) & 0 \\
0 & 0 & 0 & D
\end{array} \right)
\end{equation}
where 
\begin{equation}
D=\kappa_{S}\zeta_{S}^2
\left( \begin{array}{ccccccc}
4(\frac{1-R}{3}) & 0 & 0 & 0 & 0 & 0 & 0 \\
0 &4(\frac{1-R}{3}) & 0 & 0 & 0 & 0 & 0 \\
0 & 0 & 4(\frac{1-R}{3}) & 0 & 0 & 0 & 0 \\
0 & 0 & 0 & 2(\frac{2+R}{3}) & 0 & 0 & 0 \\
0 & 0 & 0 & 0 & 2(\frac{2+R}{3}) & 0 & 0 \\
0 & 0 & 0 & 0 & 0 & 2(\frac{2+R}{3}) & 0 \\
0 & 0 & 0 & 0 & 0 & 0 & 2(\frac{2+R}{3}) 
\end{array} \right)
\label{Hmassmat}
\end{equation}
and where an $\ast$ denotes the corresponding symmetric entry.
The rows and columns of this matrix are labelled by the Higgs
states listed in (\ref{VK}) except for 
the third and fourth row (column) which correspond respectively to 
$H_\pm = (H_2 \pm H_3)/\sqrt{2}$.

We note that apart from the $h_W$, which of course stands by itself,
the others, $H_K$, fall into 3 categories, with masses proportional
respectively to $\sqrt{1+2R}, \sqrt{1-R}, \sqrt{2+R}$.  A few of
them are mixed in the present basis, but most of them are already
diagonal.  Among the 9 $H_K$'s, there are 4 with masses proportional
$\sqrt{1-R}$.  Now recall from Figures \ref{Rplot} 
and \ref{thetaplot} the behaviour of
$R$ as the scale $\mu$ decreases from $\infty$.  By the scale of
around 20 MeV when $\theta$ crosses $0$, $R \rightarrow 1$ so that
the states $H_K$ with masses proportional to $\sqrt{1-R}$ approach
zero also.  In other words, in the present fit, there will indeed 
always be $H_K$'s with low enough masses to drive rotation down to 
the low scales one wants.

The other $H_K$ states with masses proportional to $\sqrt{1+2R}$ 
and $\sqrt{2+R}$, however, will remain of order $\zeta_S$, 
whose magnitude one has at present no estimate for, 
but presumably $> {\rm MeV}$.  In that case, they 
cannot, according to the above, contribute to driving the rotation 
at very low scales, so that the RGE derived in Section 3 by summing
over all $H_K$ should in principle be modified.  Unfortunately,
to take account of the suppression of these states with its many 
inherent uncertainties, would introduce more freedom and parameters 
than can be handled at present.  These modifications, however, will
likely affect only the rotation speed of $\balpha$ in (\ref{Rdot}) 
and (\ref{thetadot}), not the shape of $\Gamma$ in (\ref{shape}) which
comes from the residual
$\widetilde{su(2)}$ symmetry mentioned in (A).

\vspace{.5cm}

[II] To help gauge the significance of the fit given in Section 4,
one could compare it with other fits of the same data, but we are 
not aware of other fits in the literature attempting, all at one 
go, a similar fit to the mass and mixing data for both quarks and 
leptons, excepting some of our own.  Previous to the present, 
our most successful fit to these data was the phenomenological fit of
\cite{compmec}, to which 
one can thus make comparison. 

Such a comparison, however, would be grossly unfair, given that in 
\cite{compmec} one is allowed to choose freely both the shape and 
speed functions for the rotation trajectory so as to fit the 
data, whereas here, in the present fit, both these functions are 
constrained by the RGE derived from the FSM.  Nevertheless, the 
comparison in quality of the resulting fits is very much in favour 
of the present one.  For scales $\mu$ above 20 MeV, the fits are similar 
in quality, as can be seen by comparing Table \ref{muitable} to 
corresponding results in \cite{compmec}, the small differences being 
due just to differences in emphasis on which bits of data one chooses 
to fit better than others.  For scales below 20 MeV, on the other 
hand, the present fit wins on 3 significant points: 
\begin{itemize}
\item $m_e$ is on the dot, 
\item $m_u < m_d$, 
\item the trajectory has finite length, thus avoiding there being 
too many solutions for the the lowest generation,
\end{itemize}
none of which is matched by \cite{compmec}.  Although the first 
result can be ascribed to an extra parameter in the present fit 
(7 here instead of 6 in \cite{compmec}), the other 2 results 
are ``predictions'' of the FSM, which was not, and could not have 
been, anticipated in the purely phenomenological approach taken
in \cite{compmec}.  

\vspace{.5cm}

[III] Besides fitting existing data as is done in this paper, one
could try to test the FSM by predicting some new phenomena to be 
tested by experiment.  However, for these, one will need first 
to develop logically the rules in FSM for calculating quantities 
such as decay widths and scattering amplitudes and so on, i.e., beyond 
the single particle quantities like masses and mixing parameters 
considered in this paper.  To develop such rules will take time, 
given the new concept of a rotating mass matrix involved. 

In the absence of standard rules to perform actual calculations,
one can at present only make surmises, but some of these might already
be
quite exciting.  For instance, an examination of Higgs decays in
\cite{r2m2,Hdecay} led to the tentative conclusion that there may
be, in these decays, sizeable flavour-violating modes, based on  
arguments which ran briefly as follows.  Recall first the Yukawa 
coupling in (\ref{wYukawa}) above from which the quark and lepton 
mass matrix (\ref{mfact}) was derived.  From this it would seem to 
follow that the coupling of the Higgs boson would carry with it 
the same factor $\balpha \balpha^\dagger$ as did the mass matrix.
Since $\balpha$ is supposed to rotate with scale, one has to
specify at which scale to take this $\balpha$, and we suggested
$\mu = m_H$.  Sandwiching this now between, say, a lepton state 
$\ell$ on one side and another lepton state $\ell'$ on the other, 
one would obtain an amplitude for $H \rightarrow \bar{\ell} \ell'$ 
proportional to $\langle \ell|\balpha \rangle \langle \balpha|
\ell' \rangle$, or a decay width proportional to $|\langle \balpha|
\ell \rangle|^2 |\langle \balpha|\ell' \rangle|^2 $. which has
no identifiable reason to be zero for $\ell \neq \ell'$.  Hence 
flavour violation as anticipated.

Given the above fit reported in Section 4, $\balpha$ is known
at $\mu = m_H$, now measured experimentally at 126 GeV, and since 
the state vectors $\btau$ and $\bmu$ are also known, the actual
width of the flavour-violating mode $H \rightarrow \bar{\tau} \mu$, 
say, relative to $H \rightarrow \bar{\tau} \tau$ can be estimated:
\begin{equation}
 \frac{\Gamma(H \rightarrow \bar{\tau}\mu)}
     {\Gamma(H \rightarrow \bar{\tau}\tau)} 
= \frac{|\langle \balpha|\mu \rangle|^2}
       {|\langle \balpha|\tau \rangle|^2} 
\sim 2 \times 10^{-4}.
\label{BRmutau}
\end{equation}
This is still some 2 orders below the bound given recently by 
CMS for this mode, but in future might be an interesting point 
to watch.\footnote{Intriguingly, CMS \cite{CMS} actually saw 
a slight excess above background for this mode, but only at the
$2 \sigma$ level.  Besides, with a best-fit BR of about 0.009,
this excess would in any case be much too large to be explained
by the above effect.}

\vspace{.5cm}

[IV] There is yet a wide area of phenomenology opened up by the
framing hypothesis waiting to be explored.  The new ingredients
added by framing to the standard model are the framons, of which
the weak component is identified as the standard Higgs field.  This
leaves the strong framons as the main new ingredients, and it is
these that give rise, according to the RGE derived in Section 3, 
to the rotation of the mass matrix, and hence to the fermion mass
hierarchy and mixing phenomenon, the effects which gave the first
incentive for the framing hypothesis.  But now, given these new 
framon degrees of freedom, would it not also lead to new physical 
phenomena that one has not anticipated?  And may not these afford 
even better and more direct tests of the framing hypothesis than 
the rather round-about test that we have here performed?  Would 
it not be more direct experimentally to try looking for these 
strong framons themselves?  

The strong framons, however, are coloured, and so are confined by 
colour $su(3)$.  Hence, they will not exist as free particles in 
space but have to be looked for inside hadronic matter.  There, 
they can contribute to renormalization effects, for example, as
gluons do, to the running of the strong coupling $g$.  But here, 
being scalar bosons, they contribute very little to the running
speed, which is unlikely to be noticeable for some time
\cite{alpharunning}.  
There may, of course, 
be other effects in which the renormalization by strong framon 
loops is more prominent.  The rotation of the fermion mass matrix 
investigated above is one such example, but we do not yet know of
others.  

One might, perhaps, expect to see framons hadronizing, as quarks 
and gluons do, and appearing as jets in collisions with high impact.  
But this seems to us unlikely, for strong framons in (\ref{VPhi}) 
have $\mu_S > 0$, which means that they are ``ghosts'' with 
imaginary masses, and hence, unlike quarks and gluons, cannot  
propagate in hadronic matter.

However, a strong framon can form bound states with a strong 
antiframon by colour confinement, thus $\bar{\bphi}^{\tilde{a}} 
\bphi^{\tilde{b}}$, which are then the strong Higgs states $H_K$, 
the loops of which generate the RGE's in Section 3, and 
the tree-level mass matrix of which is given in (\ref{Hmassmat}) 
above.  These being just hadrons, can propagate freely in space
and be detected by experiment in principle.  Unfortunately, one 
does not yet know exactly where to look, because their masses 
depend on parameters such as $\zeta_S$, $\kappa_S$  
and $\lambda_S$ for the 
values of which one has no estimate at present.  Suppose these 
parameters are such that all $H_K$'s have large masses, then, 
being hadrons, they are likely to have big widths as well, in 
which case they will be hard to find and might escape detection 
up until the present.  But in that case, it would be hard to 
understand why they can still drive rotation of the fermion mass 
matrix down to the scales we want.  But, on the other hand, if 
we accept the conclusion in [I] above that some of the the $H_K$'s 
have small masses at low scales, low enough to drive rotation 
to the low scales we want, then they should be observable.  But 
then where are they, and why have they not been seen?  There 
is, intriguingly, an exciting possible solution to this apparent
dilemma which we have been considering recently, but this being 
rather speculative at present, even if it should work out, would 
be at variance with the stolid pragmatism intended for the present 
paper.  We shall leave it, we hope, to be presented elsewhere.

\vspace{1cm}

\appendix

\begin{flushleft}

{\bf \Large {Appendix.  Some details to facilitate spot checking of 
numerical results}}

\end{flushleft}

\vspace{.5cm}

In solving the equations (\ref{Rdot}) and (\ref{thetadot}) numerically
by iteration, a technical point is encountered which is worth noting.  
These equations are given in polar co-ordinates $\theta, \phi$ where 
by standard convention, $0 < \theta < \pi,\ 0 < \phi <2 \pi$, so 
that, given the sign in (\ref{thetadot}), $\theta$ will decrease with 
increasing $\mu$.  On reaching the point $\theta = 0$, therefore, and 
also thereafter when $\mu$ increases further, the output $\theta$ from
iterating the equation (\ref{thetadot}) must turn negative, just by 
continuity as befits a solution of differential equations.  This may 
seem disturbing at first sight, but is in fact no more than a renaming 
of the points further along the trajectory beyond $\theta = 0$ by the 
unconventional, though equally unambiguous, polar co-ordinates $-\theta, 
\phi$ instead of the more conventional $\theta, \pi + \phi$.  In the 
equations themselves, however, the $\theta$ appearing there is still
to be understood in the standard convention so that the output $\theta$
will continue to decrease with increasing $\mu$ and become more and 
more negative.  One sees then that when thus understood, the solution
will just go smoothly over $\theta = 0$, and continue on to the other 
side.  What we actually did, as is natural in FSM, however, was the 
other way round, i.e., iterate from high to low scale instead.  We then
found it convenient to adopt the convention that the output $\theta$ 
is negative at high scales, then changes sign to positive when $\theta$ crosses 
0 into the low $\mu$ region.  This is the convention adopted in the
value for $\theta_I$ in (\ref{params}) and in Figure \ref{thetaplot}. 

For the parameters given in (\ref{params}), our numerical solution
of the RGE gives $\balpha$ at the mass scales of the various quark
and lepton states as:
\begin{eqnarray}
\balpha^\dagger (\mu = m_t) & = & 
            (-0.89580, \ \  0.37467, \ \ 0.23909 )  \nonumber \\
\balpha^\dagger (\mu = m_b) & = &  
            (-0.90212, \ \  0.34069, \ \  0.26479)  \nonumber \\
\balpha^\dagger (\mu = m_\tau) & = & 
            (-0.90435, \ \  0.31923, \ \  0.28329)  \nonumber \\
\balpha^\dagger (\mu = m_{\nu_3}) & = & 
            (-0.62012, \ \  0.07942, \ \  0.78048)  \nonumber \\
\balpha^\dagger (\mu = m_c) & = & 
            (-0.90487, \ \  0.30860, \ \  0.29321)  \nonumber \\
\balpha^\dagger (\mu = m_s) & = & 
            (-0.88316, \ \  0.21072, \ \  0.41907)  \nonumber \\
\balpha^\dagger (\mu = m_\mu) & = & 
            (-0.85897, \ \  0.17903, \ \  0.47971)  \nonumber \\
\balpha^\dagger (\mu = m_{\nu_2}) & = & 
            (0.17880, \ \   -0.01812, \ \   0.98372)  \nonumber \\
\balpha^\dagger (\mu = m_u) & = & 
            (0.71377, \ \   -0.69284, \ \   0.10251)  \nonumber \\
\balpha^\dagger (\mu = m_d) & = & 
            (0.80422, \ \   -0.57797, \ \  0.13847 )  \nonumber \\
\balpha^\dagger (\mu = m_e) & = & 
            (0.83746, \ \   -0.52272, \ \   0.15944)  \nonumber \\
\balpha^\dagger (\mu = m_{\nu_1}) & = & 
            (0.90360, \ \   -0.33147, \ \   0.27136)
\label{balphamu}
\end{eqnarray}
and the state vectors of the various quark and lepton states as:
\begin{eqnarray}
\bt^\dagger & = & 
           (-0.89580, \ \  0.37467, \ \ 0.23909 )   \nonumber \\
\bb^\dagger & = & 
           (-0.90212, \ \  0.34069, \ \  0.26479)   \nonumber \\
\btau^\dagger & = &
           (-0.90435, \ \  0.31923, \ \  0.28329)   \nonumber \\
\bnu_3^\dagger & = &
            (-0.62012, \ \  0.07942, \ \  0.78048)  \nonumber \\
\bc^\dagger & = &
            (-0.14421, \ \   -0.75386, \ \  0.64102)  \nonumber \\
\bs^\dagger & = &
            (0.00217, \ \   -0.61007, \ \   0.79235)  \nonumber \\
\bmu^\dagger & = &
            (0.07434, \ \   -0.53580, \ \   0.84107)  \nonumber \\
\bnu_2^\dagger & = &
            (0.77494, \ \   -0.09291, \ \   0.62517)  \nonumber \\
\bu^\dagger & = &
            (0.42041, \ \   0.53974, \ \  0.72934)  \nonumber \\
\bd^\dagger & = &
            (0.43148, \ \   0.71537, \ \  0.54961 )  \nonumber \\
\be^\dagger & = &
            (0.42028, \ \   0.78168, \ \  0.46082)  \nonumber \\
\bnu_1^\dagger & = &
           (0.12217, \ \   0.99250, \ \  -0.00393).
\label{statevec}
\end{eqnarray}

The $\balpha$'s in (\ref{balphamu}) are easily checked to lie
on the curve $\Gamma$ given by (\ref{shape}) but to check that 
they have indeed the locations on $\Gamma$ as are given, one 
will need to solve the RGE's (\ref{Rdot}) and (\ref{thetadot}) 
for the rotation speed. Next, the state vectors for the quarks 
and leptons given in (\ref{statevec}) are also easily checked
to be consistent with their definition in (\ref{Utriad}) (and
similar expressions for the other species) in terms of the 
$\balpha$'s given in (\ref{balphamu}).  From  (\ref{statevec})
and (\ref{balphamu}), then, the masses of the various quark and
lepton states can be readily calculated by (\ref{hiermass}). They will
be seen to
tally with the results given in Table \ref{muitable} and
(\ref{5others}), except for $u$, and to a less extent for $e$, 
where the agreement may not be to the accuracy indicated
for a rather 
trivial technical reason.\footnote{The left-hand side of the 
equation (\ref{hiermass})
for $m_u$ being rapidly varying as a function of $m_u$ in the
MeV region, we chose in our numerical fitting program to 
take, as solution for $m_u$, the value at which the difference 
$\Delta$ between the left-hand and 
right-hand sides of (\ref{hiermass}) changes sign from 
$m_u$ to $m_u + \delta \mu$.  This ensures the accuracy of the 
solution for $m_u$ to be within the spacing $\delta \mu$, but 
may not give to a similar accuracy
the left-hand side of (\ref{hiermass}), or $\balpha(m_u)$.  However, this
is a valid procedure because the accurate value for $\balpha(m_u)$ is
not required.  
Similar considerations apply to the other lowest
generation states.
This same
criterion for solution was not used in the control calculation
shown in column 5 of Table 2, where it has been checked that
the two sides of (\ref{hiermass})   agree to the accuracy indicated.}
Further, from (\ref{statevec}), the PMNS matrix elements for
leptons can be calculated (assuming $\delta_{CP}$ to be zero
for leptons, as was done in Table \ref{muitable}) just as the 
inner products of the state vectors, and hence the angles 
$\theta_{ij}$ to be deduced and checked against the given
values in Table \ref{muitable}. And finally, given further 
the following:
\begin{equation}
\sin \omega_U = -0.1019, \ \ \sin \omega_D = -0.2506
\label{omegaUD}
\end{equation}
obtained in our calculation, together with the value given in
(\ref{5others}) for $\theta_{CP}$, one can calculate the CKM
matrix via (\ref{tcutilde}) and (\ref{CKMtilde}) to be checked
against Table \ref{muitable}.

The lightest generation fermions, namely $u, d, e$ and $\nu_1$,
need some extra clarification.  In contrast to the 2 heavier
generations, the equations (\ref{hiermass}) for the masses of 
these lightest states can have in general multiple 
solutions\footnote{For a detailed discussion of this fact, 
see Appendix
C of \cite{compmec}.}.  In the present FSM case, there are 3
solutions for $u, d, e$ but only 1 for $\nu_1$, which last 
therefore needs no further discussion.  Of each of $u, d, e$, 
one solution is much higher in mass than the other two, and
this we discard, since it would be unstable against decay into 
the lower two.  The two remaining solutions are very close in 
mass, differing only by order 10 keV.  As already noted above, 
around (\ref{udmass}), an approximate solution for $u$ occurs 
whenever $\balpha$ crosses the $\bt\bc$-plane, if one neglects 
some small effects, but when these small effects are taken 
into account, this solution splits into 2, one placed slightly 
above and one slightly below the $\bt\bc$-plane.  We have not 
understood the reason for this doubling of what appears to be 
but a single solution.  The numbers in Table \ref{muitable} 
are for the higher of these two solutions, but since the 
differences are so small, very similar results would be 
obtained by focusing instead on the lower, or on the average 
of the two.  Obviously, some further thinking is needed from
us to understand why there should be this doubling.

\end{document}